\documentclass[aps,prb,twocolumn,float,amsmath]{revtex4}
\usepackage[dvips]{color}
\usepackage{graphicx}
\usepackage{amsmath,bm,epsfig}
\definecolor{g-blue}{rgb}{0.83,0.95,1}
\usepackage{bm}
\usepackage{amssymb}
\usepackage{amsfonts}
\usepackage{amsmath}
 \usepackage{graphicx}

\newcommand{\eq}[1]{(\ref{#1})}
\newcommand{\Eq}[1]{Eq.\,(\ref{#1})}
\newcommand{\Eqs}[1]{Eqs.\,(\ref{#1})}
\newcommand{\Fig}[1]{Fig.\,\ref{#1}}
\newcommand{\Figs}[1]{Figs.\,\ref{#1}}
\newcommand{\Sec}[1]{Sec.\,\ref{#1}}
\newcommand{\Ref}[1]{Ref.\,\cite{#1}}
\newcommand{\Refs}[1]{Refs.\,\cite{#1}}
\def\Fbox#1{\vskip1ex\hbox to 8.5cm{\hfil\fboxsep0.3cm\fbox{%
			\parbox{8.0cm}{#1}}\hfil}\vskip1ex\noindent}  

\renewcommand{\sb}[1]{_{\text {#1}}}  
\renewcommand{\sp}[1]{^{\text {#1}}}  

\def\He4 {$^4$He~}
\newcommand{\B}[1]{{\bm{#1}}}
\newcommand{\C}[1]{{\mathcal{#1}}}    

\graphicspath{{C:/Users/Anna/Dropbox/Channel/Prague/Front/TurbulentPlug_paper/Figs/}} 

\begin{document}

\title{Dynamics of turbulent plugs in  a superfluid $^4$He channel  counterflow 
	}
\author{ A. Pomyalov}
\affiliation{Dept. of Chemical and Biological Physics, Weizmann Institute of Science, Rehovot, Israel}
\begin{abstract}
Quantum turbulence in superfluid He-4 in narrow channels often takes the form of moving localized vortex tangles. Such tangles, called turbulent plugs, also serve as building blocks of quantum turbulence in wider channels. We report on a numerical study of various aspects of the dynamics and structure of turbulent plugs in a wide range of governing parameters. The unrestricted growth of the tangle in a long channel provides a unique view on a natural tangle structure including superfluid motion at many scales. We argue that the edges of the plugs propagate as turbulent fronts, following the advection-diffusion-reaction dynamics.
This analysis shows that the dynamics of the two edges of the tangle have distinctly different nature. 
We provide an analytic solution of the equation of motion for the fronts that define their shape, velocities and effective diffusivity, and analyze these parameters for various flow conditions. 
\end{abstract}

\maketitle
\section{Introduction}
Quantum properties of liquid He become apparent\cite{Donnely,2,Vinen,NemirReport}  when it is cooled below  critical temperature $T_{\lambda}=2.17$~K. Quantized part of fluid vorticity, an inviscid superfluid, forms a quantum ground state. A gas of thermal excitations represents a viscous normal fluid with continuous vorticity. The vorticity quantization results\cite{Feynman} in creation of thin quantum
vortex lines of fixed circulation. These lines form dense tangles that interacts with the normal fluid via mutual friction force.

 When placed in a channel with a temperature gradient, two components of the superfluid He  flow in opposite directions. The superfluid flows towards the heater, while the normal fluid moves away from it. Such a setting, called thermal counterflow, has been long used to study\cite{Vinen3,PeshkovTkachenko,Tough}  properties of superfluid $^4$He components and their interaction.
  Early experiments on the thermal counterflow in $^4$He in narrow channels  found a wide variety of scenarios of the vortex tangle dynamics\cite{Vinen3,PeshkovTkachenko,Tkachenko,vanBeelen82,SchwarzRosen,Awschalom84}. 
Propagating turbulent fronts and localized vortex tangles, or turbulent plugs, were observed  in long thin glass and metal capillaries\cite{PeshkovTkachenko,Tkachenko,vanBeelen82}. Depending on conditions, these plugs were either stationary, moving in one direction or expanding  both toward and away from the heater. 

The stationary, almost homogeneous  tangles, filling the whole channel, were found in relatively wide channels \cite{Vinen3,Tough,SchwarzRosen,Awschalom84}. In this case, the local variations of the vortex line density (VLD) buildup  towards the stationary regime were considered as transient effects\cite{SchwarzRosen} and most of the attention turned to studies of the steady-state properties with VLD as the main parameter of the system.

Derivation of a set of closed equations for the description of the  quantum vortex tangle dynamics  and statistics using only its macroscopic characteristics  have been an  ultimate goal since early days of superfluid $^4$He studies. The Vinen's equation \cite{Vinen3} for the time evolution of the vortex line density $\C L$ in a homogeneous tangle served as a basis of most theoretical considerations for decades (see, for example,  \Refs{vanBeelen88, nemir95,nemir11}).  A microscopic theory by Schwarz \cite{schwarz88} introduced additional structural properties of the tangle, such as root-mean-square curvature and various anisotropy parameters, as important ingredients of the theory\cite{Lipniacki01,JoiMonjovi06,aniso}. As was pointed out by Schwarz, the arguments leading to these equations for $\C L$  apply only to the average time-dependent behavior near the steady-state, although are very often used in other situations.

For a moving tangle, a number of theories \cite{vanBeelen88,nemir11,castigone95} predicted that the plug's motion is defined by  drift as a whole with a constant velocity and a diffusion-like spreading.
It was commonly assumed that the fully developed homogeneous tangle is expanding into the laminar superfluid, having  well-defined properties, the same as for the stationary homogeneous tangle. No direct experimental or numerical evidence, supporting these assumptions, was found so far. The only numerical study of such a moving turbulent plug by Schwarz\cite{schwarz90} was carried out within an approximation that ignores non-local interactions between vortex lines and was fully focused on the conditions that allow sustaining the quantum turbulence.

Recent advances in the experimental techniques, including  flow visualization\cite{Chagovets2011,LaMantia2016,Guo-PTV}, as well as increasing computing power, renewed the interest to the spatial inhomogeneity due to presence of channel walls\cite{BaggaleyLaurie, BaggaleyLaizet13,YuiTubota15, WeiTsuVinen18,DynVLD,reply,aniso,nemir18} and  spatially-resolved investigations of the transient behavior in the thermal counterflow\cite{inhomogen17}. The  latter work  showed that the vortex tangle that eventually fills the whole channel, grows starting from a  number of remnant vortex rings. These rings first form separate localized turbulent plugs, that later merge. Remarkably, the structural properties of the large-scale tangle became homogeneous soon after the merger, while the vortex line density distribution remained inhomogeneous much longer, as reflected by very different VLD build-up patterns at different locations in the channel.

In this paper, we study the dynamical and structural properties of localized turbulent plugs in the wide range of flow conditions. Unlike previous simulations of the thermal counterflow in the channel, here the vortex tangle development in the streamwise direction is undisturbed by artificial self-interactions, caused by  periodic boundary conditions. Such conditions are routinely used to ensure the quick creation of a dense tangle that is homogeneous in the streamwise direction.  Although convenient, this approach does not allow to study the natural structure of the tangle and the local  influence of the  mean superfluid velocity on the vortex lines motion. 

The paper is organized as follows. In \Sec{s:plugDyn} we consider the vortex tangle motion as a whole and the distribution of the vortex line density in the developing tangle. We start by introducing in \Sec{ss:counterflow} important notions and parameters of the  thermal counterflow in superfluid $^4$He. Then we describe the numerical setup (\Sec{ss:setup}) and the chosen ways for the characterization of the developing vortex tangle in the channel (\Sec{ss:char}). Next, we consider the spatio-temporal evolution of the tangle vortex line density (\Sec{ss:dyn}),  while peculiarities of the transient processes are discussed in \Sec{ss:trans}.  The large-scale superfluid  motion, created inside the vortex tangle, is described in \Sec{ss:LargeScaleV}.  In \Sec{ss:c22} we  focus on the structural properties of the developed tangle.
Section \ref{s:fronts}  is devoted to the study of the VLD front dynamics and structure. First, we overview  important information from the turbulent front propagation studies, relevant for the current work (\Sec{ss:backgr}). Next, we derive an equation of motion for VLD, that describes the evolution of the tangle edges, or fronts (\Sec{ARDEQ}), show that the two tangle fronts have different nonlinearity types (\Sec{ss:Ftheta}), consider the closure for the nonlinear term (\Sec{ss:model}) and  solve the equation of front motion analytically for the front shape (\Sec{ss:sol}). Then we discuss  the parameters, that describe the front propagation: the front velocities (\Sec{ss:vf}) and the effective diffusivity (\Sec{ss:Deff}). In \Sec{s:Discussion} we summarize our findings.

\section{\label{s:plugDyn} Dynamics of a turbulent plugs}
\subsection{\label{ss:counterflow} The counterflow turbulence in the channel}
As already mentioned, at temperatures below
$T_{\lambda}=2.7$\,K,  liquid $^4$He become a superfluid.  In this state, the superfluid He of the density $\rho$  is often described\cite{Donnely, Vinen} in the framework of the two-fluid model as an interpenetrating mixture of a normal fluid with the density $ \rho\sb n$ and a superfluid component of the density $\rho\sb s$, such that $\rho\sb s+\rho\sb n = \rho$ and the components' contributions $\rho\sb s, \rho\sb n$ are strongly temperature dependent\cite{DB98}. 
 
The normal-fluid component has low viscosity and continuous vorticity, while the superfluid is inviscid and its vorticity is constrained to
 vortex-line singularities
of core radius $a_0 \approx 10^{-8}$ cm 
with fixed circulation $\kappa = h/M\approx 10^{-3}$ cm$^2$/s, where $h$ is Planck's constant and $M$ is the mass of the $^4$He atom. 
The two components are coupled by the mutual friction force. 
Under the influence of the temperature gradient applied along the channel, the normal fluid is moving away from the heater with a  mean velocity $\B V\sb n$. At the same time, the superfluid is moving towards the heater with the mean velocity $\B V\sb s$, creating a relative, or a counterflow,  velocity $\B V\sb{ns}=\B V\sb n-\B V\sb s$, proportional to the applied heat flux. The chaotic tangle of vortex lines is then generated from pre-existing remnant vortex loops due to the coupling by temperature-dependent mutual friction force. The governing parameters that define the dynamics and the structure of the tangle are, therefore, the counterflow velocity and the temperature, while the geometric constraints, such as channel dimensions, influence the inhomogeneity of the vortex tangle.
\begin{figure}[t]
	\includegraphics[scale=0.5 ]{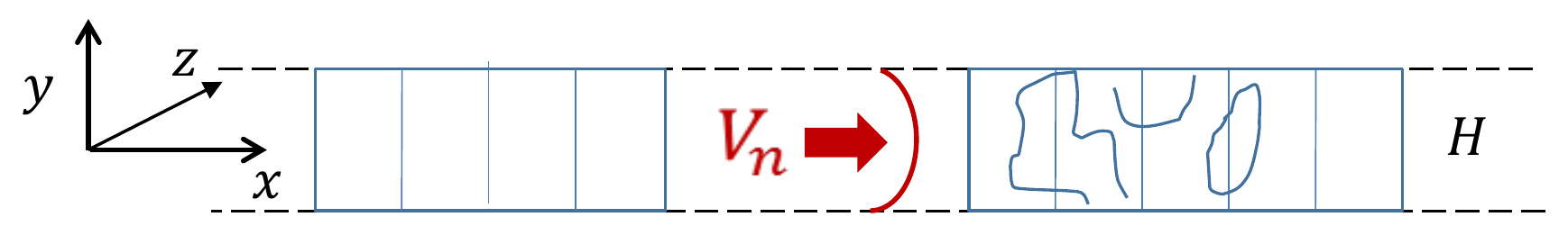}
	\caption{\label{f:1}Numerical setup. The simulations are set up in a long planar channel of a width $H$. The normal-fluid velocity is oriented towards positive $x$-direction. }
\end{figure}

\subsection{Numerical setup}\label{ss:setup}
The simulations were set up in a long  planar channel of a width $H$ (see Fig. \ref{f:1}). To describe dynamics of the vortex lines we use the vortex filament method \cite{schwarz88,recon14,Recon} for the channel flow \cite{BaggaleyLaizet13,inhomogen17,DynVLD}.  The vortex lines are parameterized  by curves $\B s(\xi,t)$ and discretized in a set of points with the  resolution $\Delta \xi=0.001$ cm. 
The equation of motion for such a line point\cite{schwarz88} is
\begin{eqnarray} \label{SFVel}
\frac{d\B s(\xi,t)}{dt} &=& \B V\sb {drift}(\B s, t)= \B V \sb s(\B s, t)+\B V\sb{mf}(\B s, t)\, ,\\\nonumber
\B V\sb{mf}(\B s), t&=&(\alpha - \alpha' \bm {s}'\times\big )  \mathbf{s}' \times \B V\sb{ns}(\B s,t)\,  ,
\end{eqnarray}
where $\B s^{\prime}$ is the unit vector along the vortex lines and   $\alpha, \tilde \alpha$ are the temperature-dependent mutual friction parameters\cite{DB98}.
Here the right-hand-side of \Eq{SFVel} represents the drift velocity of the tangle $\B V\sb {drift}$. The superfluid velocity 
\begin{eqnarray}
\B V\sb s&=&{\bm V}\sb{BS}+\B V^0\sb{s} \, ,\\
{\bm V}\sb{BS} (\B s,t)&=&\frac{\kappa}{4\pi}\int_{\Omega} \frac{\B s-\B{s_1}}{|\B s-\B{s_1}|^3}\times  \B {d s_1} =\B V\sb{loc}+\B V\sb{nl} \\\nonumber
\end{eqnarray}
accounts for the tangle contribution ${\bm V}\sb{BS} (\B s,t)$ and  the mean superfluid velocity $\B V^0\sb s$ that is defined by the counterflow condition of zero mass-flux.  In its turn, ${\bm V}\sb{BS}$ may be further divided into the local part, produced by the scales up to local radius of curvature $R=1/|s''|$,  $\B V\sb{loc}=\beta (\B s '\times \B s''), \beta=(\kappa/4\pi)\ln(R/a_0)$ and the nonlocal velocity $\B V\sb{nl}$ which is produced by the rest of the tangle $\Omega$.  The mutual friction part $\B V\sb{mf}(\B s,t)$ describes the interaction with the normal fluid via counterflow velocity $ \B V\sb{ns}(\B s,t)=\B V\sb{n}-\B V\sb s(\B s,t)$.  The material parameters of\,  $^4$He, used in the simulations, are listed in Table \ref{t:1}.
The time resolution for the vortex line point is set by the forth-order Runge-Kutta stability criterion.  
\begin{figure}[t]
	\includegraphics[scale=0.6]{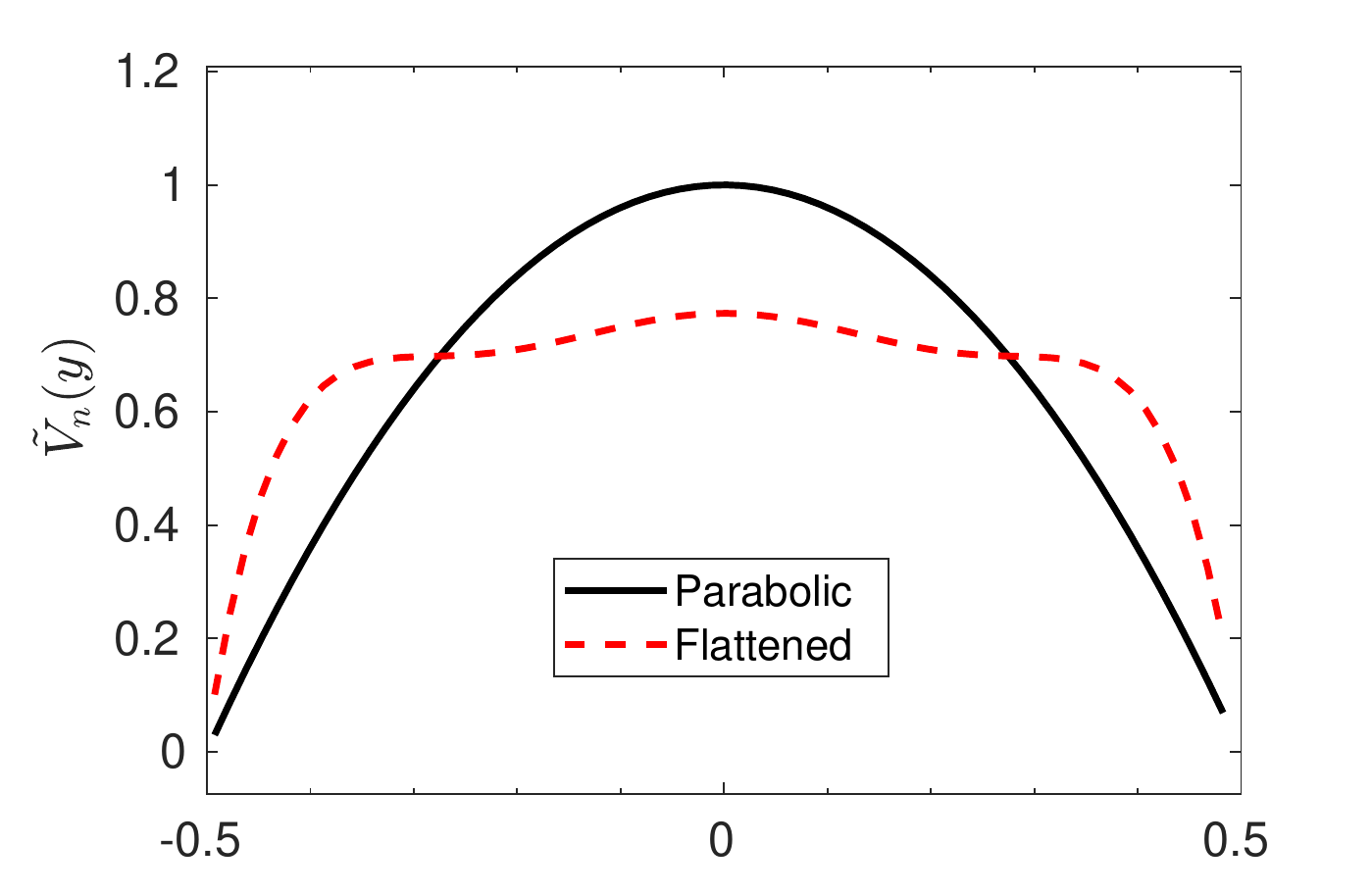}
	\caption{\label{f:2} Normal-fluid velocity profiles normalized by the mean value 
		$\tilde V\sb n=V\sb n/\langle V\sb n \rangle$. The shape of the flattened profile is defined by a combination of six Legendre polynomials, such that it has the same $\langle V\sb n \rangle$ as the corresponding parabolic profile. }
\end{figure}

To generate the counterflow, we use two time-independent prescribed wall-normal profiles of the stream-wise projection of the normal-fluid velocity $V^x\sb n(y)$, shown in \Fig{f:2}.  The parabolic profile corresponds to the laminar normal-fluid velocity. It was observed experimentally at low heat fluxes. At larger heat fluxes, when the normal fluid loses its stability but not yet become fully turbulent, its profile flattens \cite{Marakov15}. Similar flattening of the normal-fluid velocity profile was found in simulations with a two-way coupling of the fluid components
\cite{coupled17,TsubotaCoupled}. Although  such a fully coupled dynamics gives the most reliable description of the superfluid $^4$He, it is still computationally prohibitive for sufficiently large systems and long propagation times.  Therefore we ignore the back-influence of the superfluid component  on the normal fluid and model the expected normal-fluid velocity flattening by imposing the \\
corresponding time-independent profile (dashed line in \Fig{f:2}).

The mean superfluid velocity $V^0\sb s$ is dynamically defined by the zero-mass-flux condition
\begin{equation}\label{CCond}
	\rho\sb n \langle V\sb {\rm n}\rangle_v+\rho\sb{\rm s} \langle V^0\sb s\rangle_v=0\, ,
\end{equation}	
	 where $\langle ... \rangle_v $ denotes global volume averaging and $V^0\sb s$ include a contribution of the superfluid velocity induced by the vortex tangle, calculated on a dense grid. This contribution, although small,  is not negligible and grows with the development of the spatially-inhomogeneous tangle.

\begin{figure*}[htp]
	\includegraphics[scale=0.47]{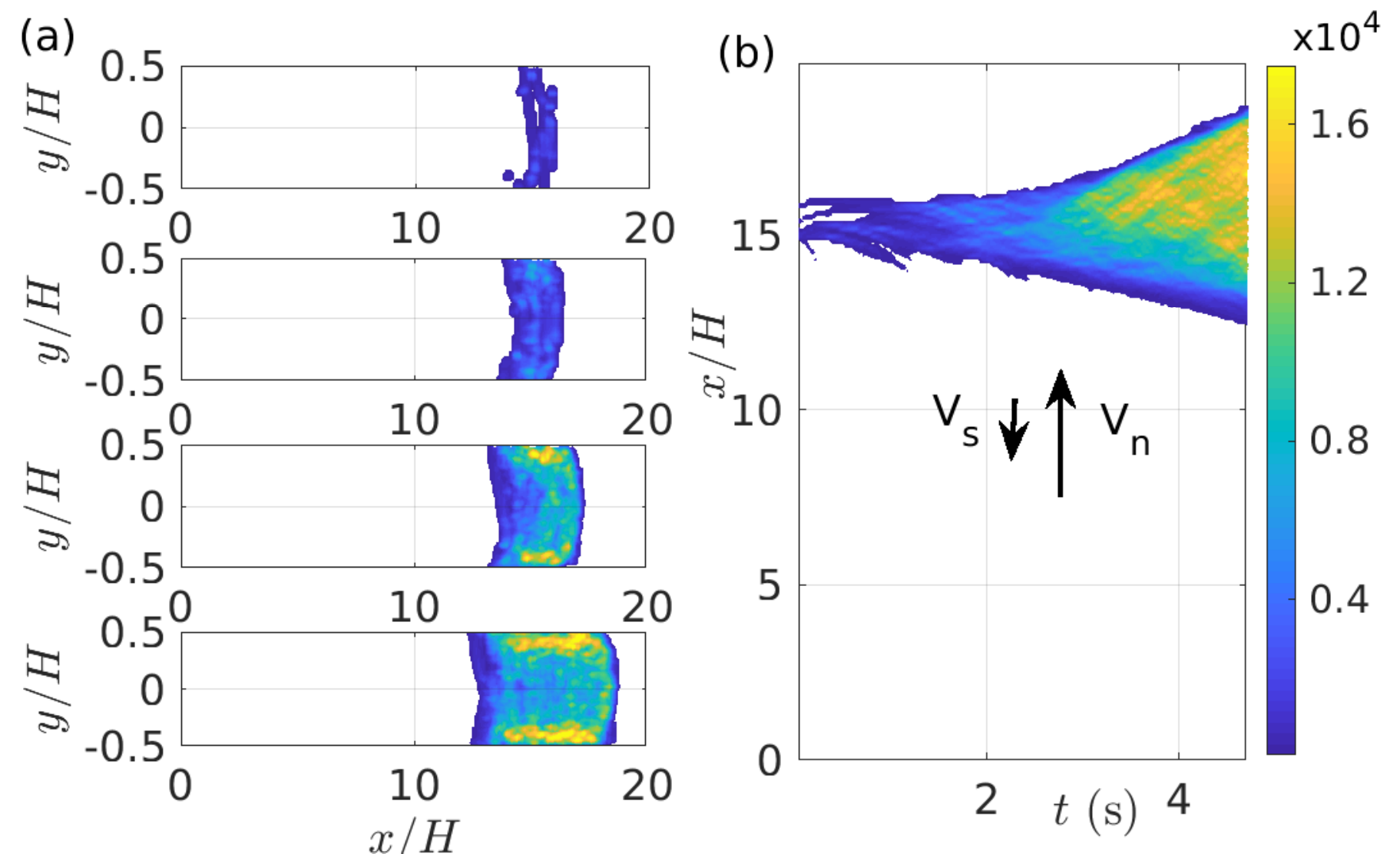}
	\includegraphics[scale=0.47]{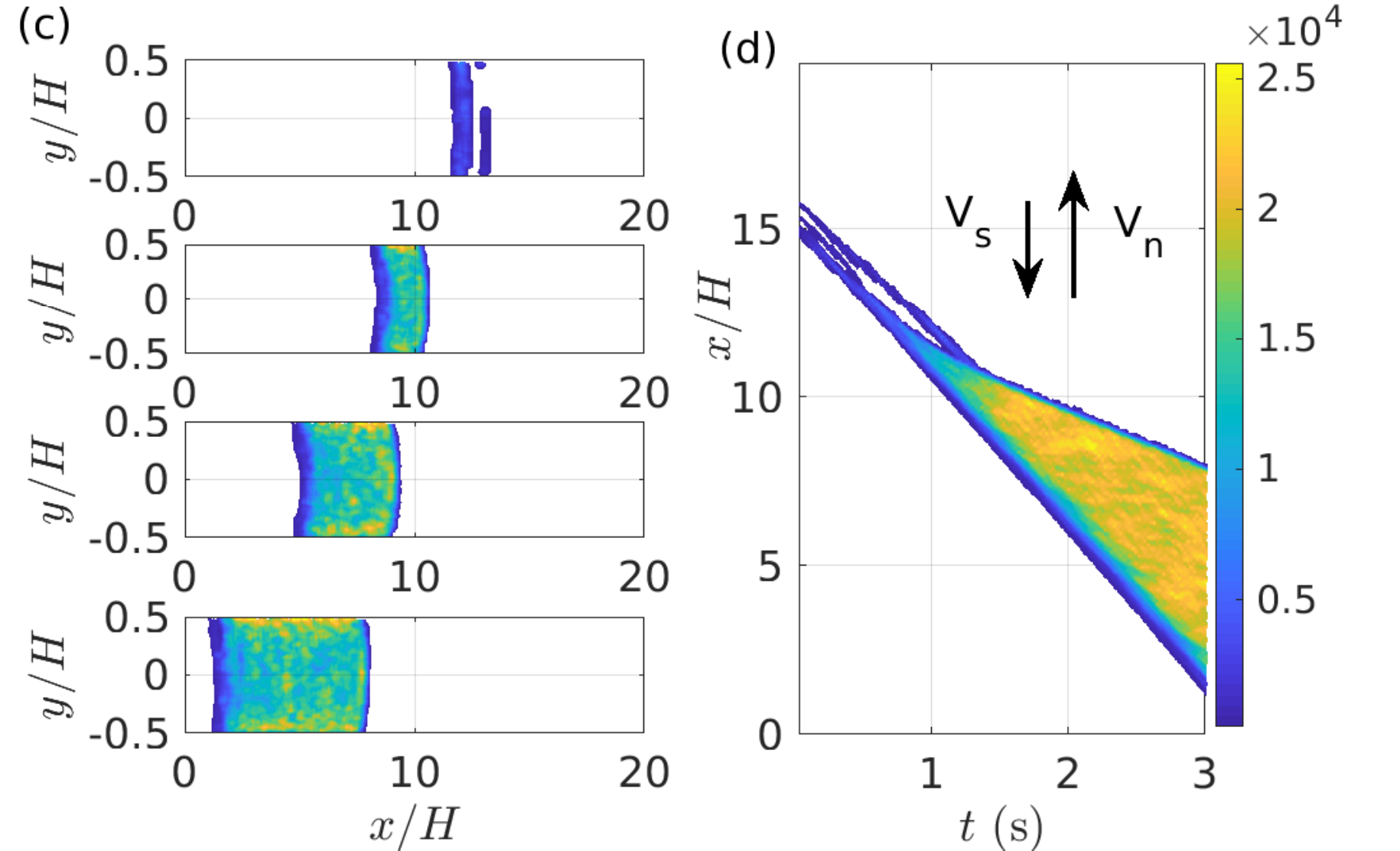}
	\caption{\label{f:3}VLD evolution. (a)-(b)  $T=1.3$ K, $U_c=3$ cm/c. (c)-(d)-  $T=1.9$ K, $U_c=1$ cm/c. Panels (a) and (c) show $ {\C L}(x,y)$ distribution at $t=0.2\,$s, $  t\sb f/2, 3 t\sb f/4$ and $t\sb f$  with the top snapshot corresponding to the early stages of the dynamics and the bottom snapshot corresponding to $t\sb f$. Panels (b) and (d) show the time evolution of VLD averaged over $y$ direction ${\C L}(x,t)$. Both cases correspond to the parabolic profile of $V\sb n$ and the channel width $H=0.1$ cm. The values of $\C L$ are color-coded as shown by colorbars in panels (b) and (d).}
\end{figure*} 

To mimic solid walls in the wall-normal direction, the boundary conditions on the wall are $s'(\pm H/2)=(0,\pm 1)$  and $V^y\sb s(\pm H/2)=0$. In the spanwise direction, periodic conditions  were used. 
To ensure  free evolution of the developing tangle, we use  open conditions in the streamwise direction. In this way, the properties of the tangle edges, moving as fronts, as well as the natural structure of the tangle bulk, can be studied.

The vortex tangles  at all conditions were initiated using the same set of 8 vortex loops of similar sizes $R_0\ll H$ and different orientations. The loops were placed at a particular streamwise location, 4 circular loops in the bulk and 4 half-circular loops at the walls. The difference in the dynamics of these tangles, therefore,  originates from the flow conditions only, allowing comparison. 

\begin{table}[t]
	\caption{\label{t:1}  Material properties\cite{DB98} of $^4$He used in the simulations.
	 }
\begin{tabular}{   c|c c c }
		\hline
	       $T$,  K    ~~~         	&~~~~1.3~~~~ & ~~~~1.65~~~~&~~~1.9~~~~~~  \\ \hline
	$\rho\sb n/\rho\sb s$~~~&   0.047      &   0.239       & 0.723   \\
	$\alpha$~~~ &   0.034      &  0.11          &0.206\\
			$\alpha'$ ~~~&    0.0138      & 0.0144           &0.0083	\\
	\hline
\end{tabular}
\end{table}

The tangle dynamics was studied at three temperatures $T=1.3, 1.65$ and $1.9$ K.
Other simulation parameters include various normal-fluid velocities.
In most simulations, the parabolic profile for $V\sb n$ and a narrow channel width $H=0.1$ cm was used. At each temperature, one case 
was chosen for simulations  with wider channels
and with flattened normal-fluid velocity profile (the same $\langle V\sb n \rangle$ as for the corresponding parabolic profile).  The simulation parameters are listed in Table~\ref{t:2}, columns $\#2-\#7$. In all simulations, the size of the channel  in $z$-direction was always equal to $H$. Despite the periodic boundary conditions in the spanwise direction,  the interaction between the vortex lines and their images is an important factor in the current setting. The study of the influence of the slit aspect ratio on the tangle dynamics is beyond the scope of this paper. The tangle evolution was followed until a well-developed bulk tangle  was formed, such that the final length of the tangle is about $4-8 H$. The actual final time of evolution $t\sb f$ varies for different conditions.

\begin{table*}[t]
	\caption{\label{t:2}  Parameters of simulations by columns:
		(\#\,1) Run \#, (\# 2) Temperature  (K); (\# 3) Type of $V\sb n$ profile: P denote  parabolic  profile, F denote for flattened profile; (\# 4)  Channel width;  (\# 5)   Centerline velocity $U_c$; (\# 6) Mean normal-fluid  velocity $\langle V\sb n\rangle$. For the parabolic profile, $\langle V\sb n\rangle=-2/3\,U_c$; (\# 7) Mean counterflow velocity $V^0\sb{ns}=\langle V\sb n \rangle_y (1+\rho\sb n/\rho \sb s)$;  (\# 8)  Bulk VLD in the core of the channel $\C L\sp {core}_0$; (\# 9)  Bulk VLD near the walls $\C L\sp {wall}_0$. The error-bars for $\C L_0^j$ account for the standard deviation from the mean values.
		 }.
	\begin{tabular*}{\linewidth}{@{\extracolsep{\fill} }    c c c c c c c c c c  l}
		\hline\hline
		1   & 2      & 3    & 4    & 5     & 6           & 7            & 8                           & 9                           & \\ \hline
		Run & $T$, K & Type & $H$  & $U_c$ & $ \langle V\sb n\rangle $ & $V^0\sb{ns}$ & $\C L\sp {core}_0  \times 10^{-4}$          & $\C L\sp{wall}_0 \times 10^{-4}$           & \\
		\#  & K      & -    & cm   & cm/s  & cm/s                      & cm/s         &  cm$^{-2}$ &  cm$^{-2}$ & \\ \hline
		1   &      & P    & 0.1  & 2     & 1.66                      & 1.4          & $0.37\pm0.03 $     & $0.6 \pm0.1 $   \\
		2   &    & P    & 0.1  & 3     & 2                         & 2.11         & $0.95 \pm0.05 $  & $1.5 \pm0.4 $ \\
		3   &        & P    & 0.1  & 4     & 2.66                      & 2.84         & $2.20\pm0.06 $      & $3.2\pm 0.8 $   \\
		4   &  1.3~        & F    & 0.1  & $-$   & 2                         & 2.11         & $1.3\pm 0.3 $    & $1.5\pm0.2$   \\
		5   &        & P    & 0.15 & 3     & 2                         & 2.12         & $1.0\pm 0.2$    & $1.4\pm 0.4$    \\
		6   &        & P    & 0.2  & 3     & 2                         & 2.11         & $7.1\pm 0.2 $        & $8.2\pm0.2$                \\ \hline
		7   &   & P    & 0.1  & 1.5   & 1                      &    1.22      &    $0.86\pm0.02$    &        $1.3\pm0.3$       \\
        8 & 1.65 & P    & 0.1  & 2     & 1.66                     &   1.63      &    $ 1.84\pm0.03  $  &         $ 2.5\pm0.5 $           \\
		9   &        & F    & 0.1  & $-$   & 1                         &    1.2       &     $1.29\pm0.02 $   &     $1.4\pm0.2 $              \\ \hline
		10  &   & P    & 0.1  & 1     & 0.66                  &  1.19         &  $1.87\pm0.02$     &  $2.1\pm0.3$  \\
		11  &        & P    & 0.1  & 1.2   & 0.8                      &   1.42       &  $2.92\pm0.03$    &  $  3.2\pm 0.3 $   \\
		12  &     1.9     & P    & 0.1  & 1.5   & 1                         &    1.36      &    $4.6\pm0.1$    &    $4.8\pm0.6$   \\
		13  &        & F    & 0.1  & $-$   & 0.66                     &   1.17       &   $ 2.55 \pm0.05$ &     $ 2.5 \pm0.6 $  \\
		14  &        & P    & 0.15 & 1     & 0.66                     &    1.18      &  $ 1.82\pm0.04 $    &  $  2.0\pm0.2 $   \\
		15  &        & P    & 0.2  & 1     & 0.66                      &   1.17        &  $ 1.84\pm0.06  $   &     $1.7\pm0.2$   \\ \hline
	\end{tabular*}
	
\end{table*}

 \subsection{Characterization of the tangle}\label{ss:char}
 To characterize the developing tangle, we calculate the time-dependent two-dimensional (2D) $(x,y)$-spatial distribution of tangle properties, averaged over spanwise $z$ direction, at equispaced time moments. To this end, we define a fixed grid with the a resolution  $\Delta x=0.011$cm and $\Delta y=0.0015$cm. The 2D maps of the tangle properties are calculated by integration\cite{schwarz88} over parts of the tangle $\Omega'$ that fall into a grid cell $V'=\Delta x \times  \Delta y \times H$.  In such a way we obtain the vortex line density $\C L$, the curvature of the vortex lines $\varkappa \equiv |s''|$, the mean square curvature $\langle\varkappa^2\rangle$, the ratio  $c^2_2=\langle \varkappa^2\rangle/\C L$, the local binormal $\B I_{\ell}=\langle \B s^{\prime }\times \B s^{\prime\prime}\rangle $ and its anisotropy index
  $ \B I^{\dagger}_{\ell}=\langle \B s^{\prime }\times \B s^{\prime\prime}\rangle /\langle |s^{\prime\prime}|\rangle$,
  the contributions to the tangle drift velocity, as defined by right-hand-side of \eqref{SFVel} and various terms of the balance equation, defined by \Eq{EqLx}. In the above definitions, the arguments $(x,y,t)$ were omitted for  clarity. To compare the results for different flow conditions, we use dimensionless quantities, normalized  using the mean counterflow velocity calculated from  the zero-mass-flux condition $V^0\sb{ns}=\langle V\sb n \rangle_y (1+\rho\sb n/\rho \sb s)$   and the circulation quantum $\kappa$.  The procedures for calculation of various profiles are described in Appendix \ref{a:1}.
  
  To measure the velocity of front propagation, it is customary to choose a threshold value of propagating quantity and to follow the change of its position. To avoid inevitable freedom  in the choice of the threshold value $\C L$, we use here a different approach. Instead of following a single threshold value, we find the velocity that allows to collapse whole edge of the tangle to a single shape. It turned out that such an approach gives a very robust measurement of the velocity, allowing simultaneously to study the front shape.
  The speeds of both VLD fronts were measured over the time interval when the tangle bulk is formed and the fronts do not change their shape during propagation.  The details on the procedure are described in Appendix \ref{a:2}.  The values of bulk VLD in the channel core and near the walls are listed in Table \ref{t:2}, columns $\#8-9$. The error-bars here and in \Figs{f:c22all}, \ref{f:Bdag}, correspond to the standard deviation around the mean values. 

  \subsection{\label{ss:dyn}Evolution of VLD}
 
 The examples of the evolution of the vortex line density at low and high temperatures are shown in \Fig{f:3}. These examples illustrate the main difference in the flow conditions that crucially affect the tangle dynamics.   The vortex  tangle is advected by the superfluid velocity field. At low $T$, the mean superfluid velocity $V^0\sb s$ is weak due to small the fraction of the normal fluid [cf. \Eq{CCond}]. The tangle dynamics is governed mostly by the tangle-induced velocity and a net tangle displacement is negligible, as is illustrated in \Fig{f:3}b. On the other hand, at high temperature, $V^0\sb s$ and $V\sb n$ are comparable and the tangle is flushed along the channel by the mean superfluid velocity, see \Fig{f:3}d.  Under all conditions, the vortex tangle develops as a moving  turbulent plug.  At $T=1.9$~K, the initial vortex rings are at first separated into at least two groups that grow into independent turbulent plugs that later merge. The developing tangles are inhomogeneous in both the streamwise and wall-normal directions, as is illustrated by snapshots of 2D VLD distributions at various time moments in \Fig{f:3}a,c. 
  The VLD is higher near the walls, similar to the steady-state tangles with the parabolic profile of the driving normal-fluid velocity, obtained under periodic streamwise conditions\cite{BaggaleyLaurie,BaggaleyLaizet13,YuiTubota15,DynVLD,coupled17}. Two edges of the tangle are different: a narrow and sharp edge is formed in the direction of $\B V\sb n$ and a wide and less steep edge in the direction of $\B V\sb s$. As we show later, these edges move with  constant velocities and without changing their  shape. We, therefore, label them as a hot front (moving the direction of normal-fluid velocity away from the heater) and a cold front (moving in the direction of  mean  superfluid velocity toward the heater). In further analysis we distinguish a near-wall and a core regions in the wall-normal direction and a bulk and the fronts regions in the streamwise direction, see \Fig{f:diagram}. 
 
\begin{figure*}[htp]
  	\includegraphics[scale=0.6]{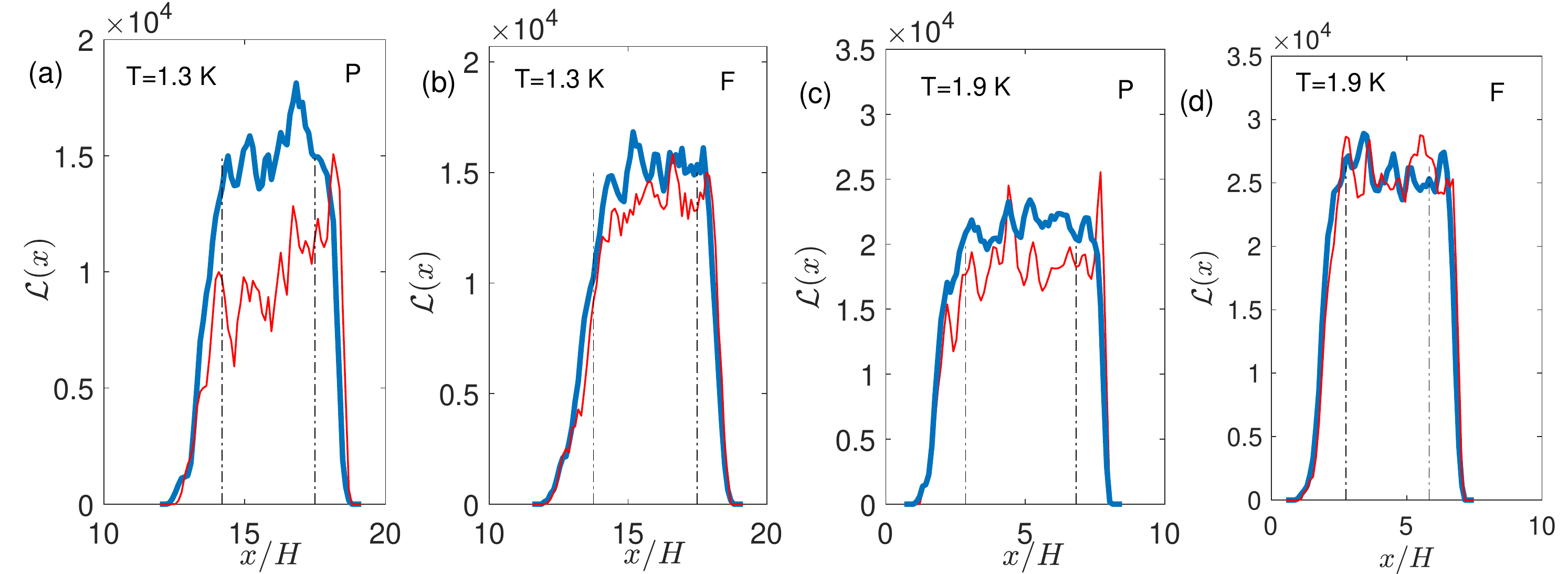}	
  	\caption{\label{f:4}The stream-wise VLD profiles  $\C L(x)$ for $T=1.3$ K [panels(a) and (b)] and   $T=1.9$ K [panels(c) and (d)]. The parabolic profile  of $V\sb n$ ( ``P") was used in panel (a) [$U_c=3$ cm/s] and panel (c) [$U_c=1$ cm/s]. In the panels (b) and (d), the flattened $V
  		\sb n$ profiles (``F")  were used, with the same $\langle V\sb n \rangle$  as in the panels (a) and (c), respectively.  The profiles for the walls region are shown by  thick blue lines and for the core region by thin red lines.  Vertical dot-dashed lines  denote the edges of the bulk region.  The channel width is $H=0.1$ cm,  $t=t\sb f$. }
  \end{figure*}

\begin{figure*}[htp]
  		\includegraphics[scale=0.6]{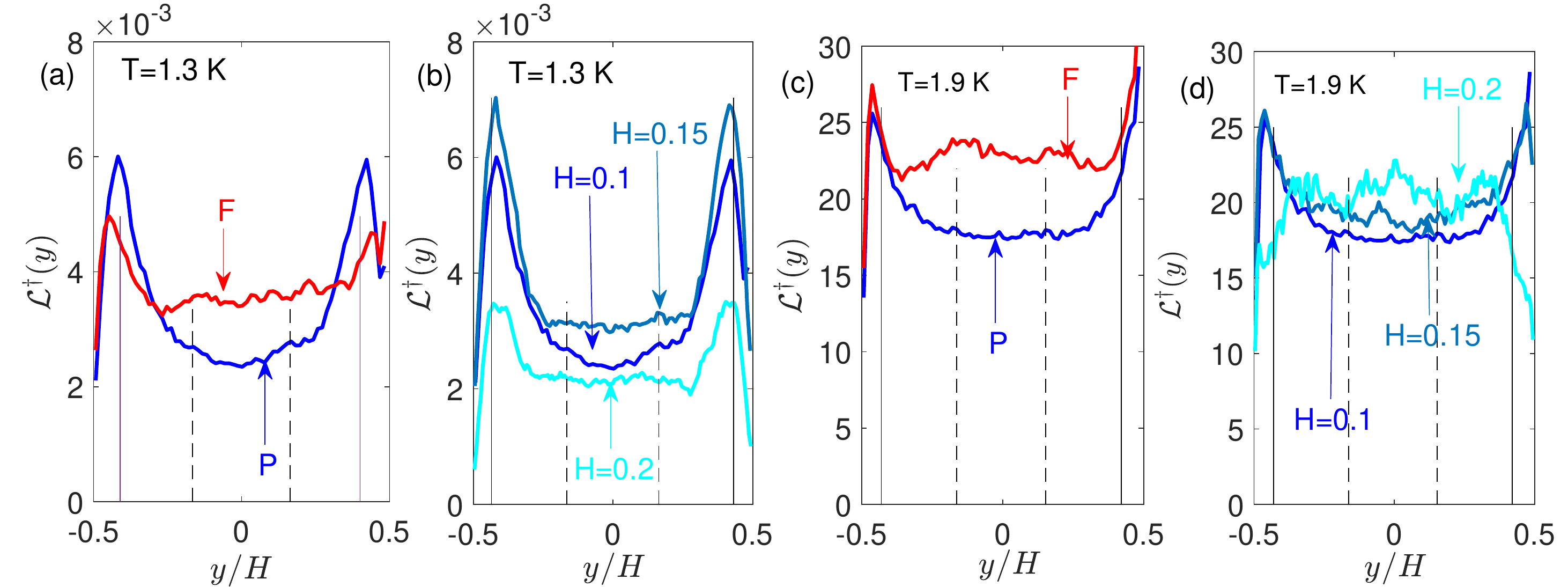}	
  	\caption{\label{f:5}The wall-normal profiles of normalized VLD profiles $\C L^{\dagger}(y)$ for $T=1.3$ K [(a) and (b)] and   $T=1.9$ K [(c) and (d)].  Panels (a) and (c) compare the $\C L^{\dagger}(y)$ profiles for the parabolic (P) and the flattened (F) $V\sb n$ profiles at $H=0.1$~cm. The panels (b) and (d) compare the $\C L^{\dagger}(y)$ profiles for the parabolic $V\sb n$ and different channel widths labeled in the figures by their values.  All profiles at $T=1.3$ K correspond ot $U_c=3$ cm/s, the profiles at $T=1.9$ K correspond to $U_c=1$ cm/s.  Here, and in \Figs{f:6}-\ref{f:8}, dashed vertical lines  denote edges  of the channel core. Thin solid black lines are placed at the intervortex distance from the walls. In each panel only one intervortex distance is shown to avoid clutter. }
  \end{figure*}

 To characterize the distribution of VLD along and across the tangle, we plot its streamwise profiles in  \Fig{f:4} and wall-normal profiles in \Fig{f:5}. The  dynamics of VLD, obtained with the parabolic normal-fluid velocity at various values of $U_c$, differ mostly by the duration of transient behavior in the tangle core and  the mean value of VLD in the bulk of the tangle.

 In \Fig{f:4} we compare the streamwise VLD profiles for the parabolic and for the flattened normal-fluid profiles at similar $t\sb f$.  At low $T=1.3$~K (\Fig{f:4}a,b), the vortex line density at the walls in both cases reached similar  values, while the core region for the parabolic $V \sb n(y)$, shown in  \Fig{f:4}a,  is still not fully developed (see \Sec{ss:trans} for details). The length of the tangle in both cases is about 3.5 $H$. The edges of the tangles reached similar streamwise positions indicating similar fronts velocities. At high $T=1.9$~K, \Fig{f:4}c,d,  $\C L(x)$ for both $V\sb n$ profiles is almost homogeneous in the tangle bulk. Here, however, the hot edge  moved faster for the flattened $V\sb n(y)$, leading to a shorter plug. The mean VLD in the bulk, in this case, is about 20\% higher than for the parabolic profile.

 To compare the wall-normal VLD profiles for various flow conditions, we plot in \Fig{f:5} a dimensionless  VLD $\C L^{\dag} (y)=\C L(y) \kappa^2/ (V^0\sb{ns})^2$. Here we compare  $\C L^{\dag}(y)$ for the parabolic and flattened $V\sb n(y)$  for the narrow channel $H=0.1$ cm in panels (a) and (c) and for various channel widths, using the parabolic normal-fluid velocity profile with the same $U_c$, in panels (b) and (d).  

 \begin{figure*}[htp]
 		\includegraphics[scale=0.6]{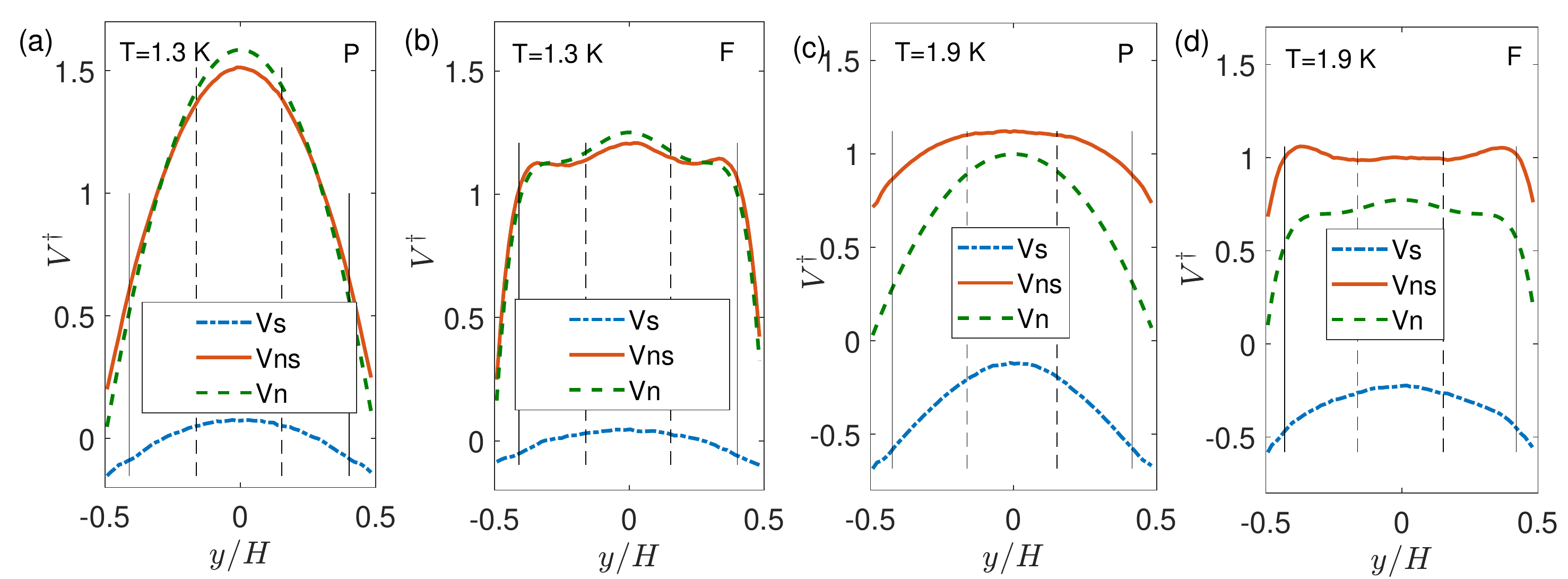}
 	\caption{\label{f:6}The wall-normal profiles of normalized velocities $V^{x \dag}\sb {s}, V^{\dag}\sb n$ and  $V^{\dag}\sb {ns}$.  Panels (a) and (b) compare the profiles for the parabolic (``P") and the flattened $V\sb n$  (``F") profiles for $T=1.3$~K, $U_c=3$ cm/c, panels (c) and (d) compare the velocity profiles  for $T=1.9$~K, $U_c=1$ cm/c. The channel width is $H=0.1$~cm. Green dashed lines denote the driving $V\sb n(y)$ profiles, blue dot-dashed lines denote the full  streamwise  superfluid velocity $V^x\sb s(y)=V^0\sb s+V^x\sb{BS}(y)$, solid red lines denote the profiles of the counterflow velocity $V\sb{ns}(y)=V\sb n(y)-V^x\sb s(y)$.
 	}
 \end{figure*} 

 First of all, we note that the wall-normal profiles $\C L(y)$ are consistent with the profiles obtained in the steady-state tangles \cite{BaggaleyLaizet13,DynVLD} with the  VLD peaking near the wall at about the intervortex distance. The flattened profiles [thin red lines in panels (a) and (c)] have larger VLD values in the tangle core. The $\C L^{\dag}(y)$ near the walls is higher for the parabolic $V\sb n$ profile at $T=1.3$~K and similar for both profiles at $T=1.9$~K.  At both temperatures, VLD for the flattened $V\sb n(y)$ is homogeneous not only over the core region but also over a large part of the near-wall region, especially at $T=1.9$~K. A similar effect of flattening of VLD profile is observed in wider channels for the parabolic $V\sb n(y)$, indicating that the increase of VLD near the walls is indeed related to the boundary effect.  These profiles are compared in \Fig{f:5}b,d. At both temperatures, the tangles for widest channels $H=0.2$ cm are not yet fully developed, although in a different way: at low $T$ the VLD just did not reach the expected values, while at high $T$ the tangle is formed by merging of two independent vortex plugs [similar to shown in \Fig{f:3}d]. The resulting streamwise inhomogeneity does not allow to properly resolve the near-wall region in the profiles calculated over narrow tangle bulk. However, it can be clearly seen that, as the channel become wider, the range of nearly-flat VLD distribution extends from the core to the near-wall region, in a way similar to the flow, generated by the flattened $V\sb n(y)$ profile. Comparing the VLD profiles for $H=0.1$ and $0.15$cm,  we find that at low temperature the VLD is peaking stronger near the walls, while at high $T$ the near-wall VLD is similar for both channel width. The normalized positions of the peaks do  not change with the channel width, meaning that the peaks appear further from the wall for wider channels.

 \begin{figure}[htp]
 	\includegraphics[scale=0.58]{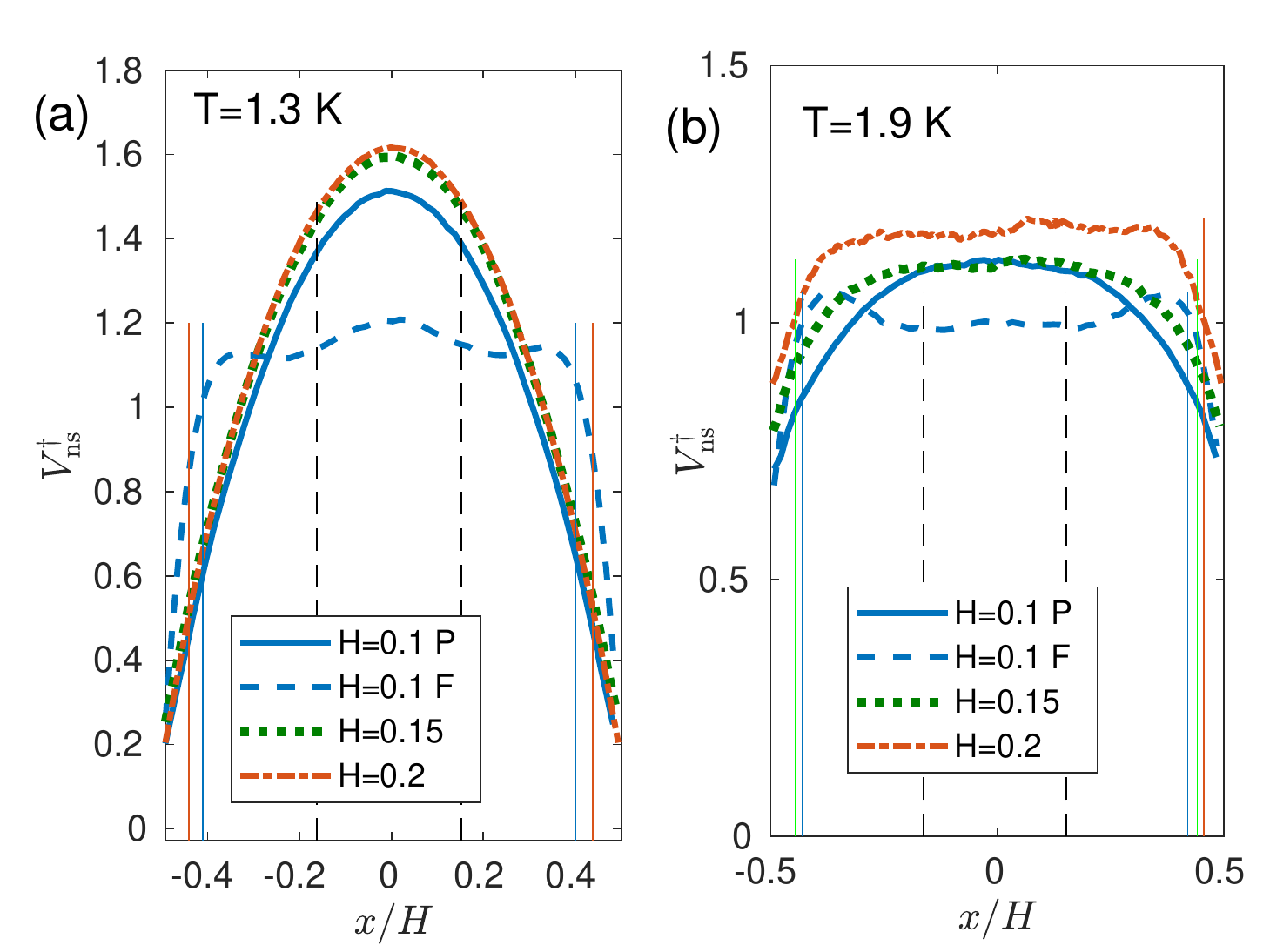}
 	\caption{\label{f:7}The wall-normal profiles of normalized $V^{\dagger}\sb {ns}$ profiles for the parabolic  at various channel widths and for the flattened $V\sb n$ (dashed lines)  profiles for (a) $T=1.3$ K  and for (b) $T=1.9$ K. The solid lines correspond to $H=0.1$ cm, dotted lines for $H=0.15$\,cm, dot-dashed lines correspond to  $H=0.2$ cm. Thin vertical lines are placed at the intervortex distance from the walls, their colors match the color of the corresponding velocity profiles.}
 \end{figure} 

 \begin{figure}[htp]
 	\includegraphics[scale=0.58]{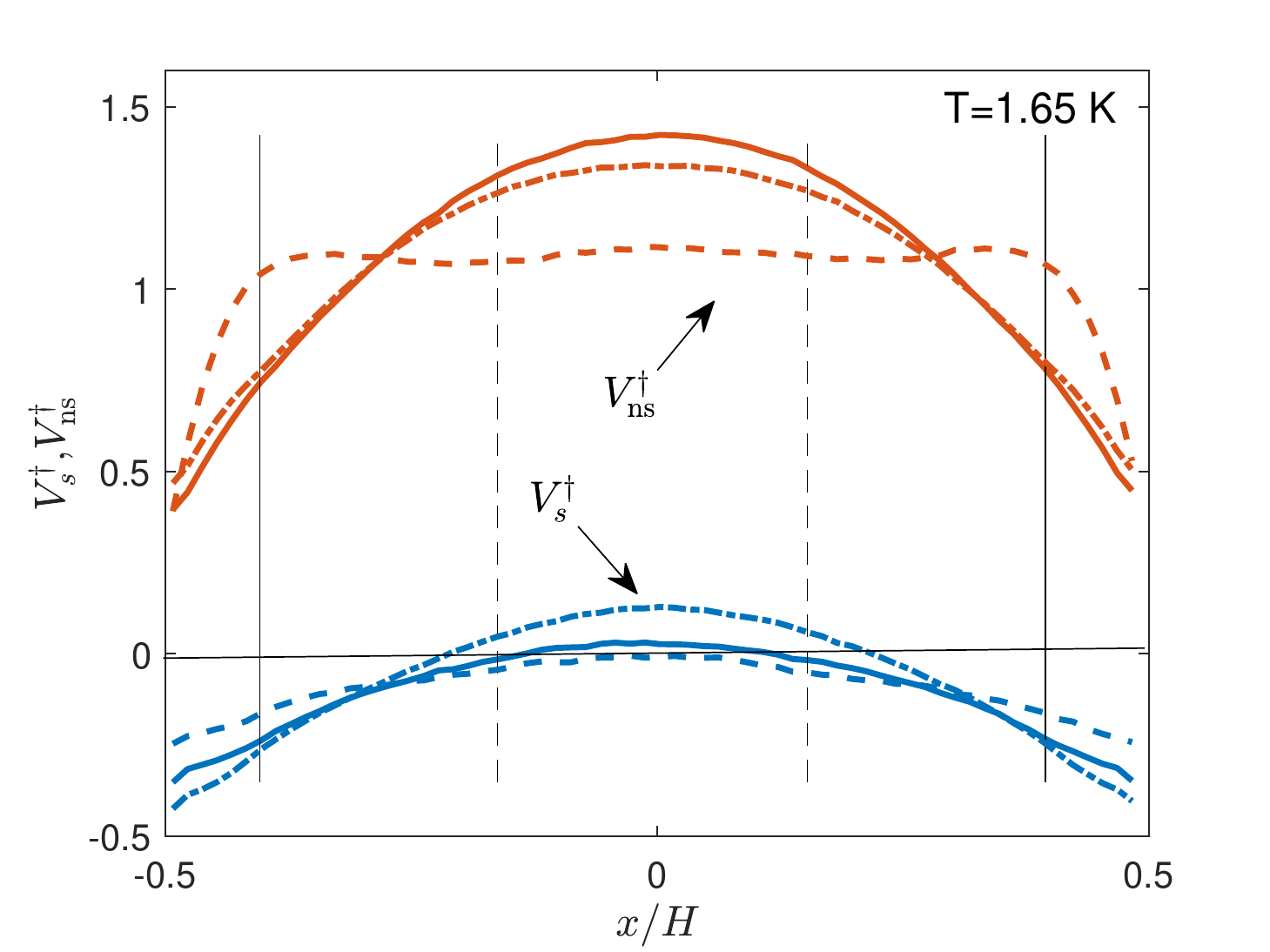}
 	\caption{\label{f:8}The wall-normal profiles of normalized  $V\sb {s}$ and $V\sb {ns}$ profiles for $T=1.65$ K and the channel width $H=0.1$cm. Solid lines correspond to the parabolic $V\sb n$ with $U_c=1.5$ cm/s, dot-dashed line to $U_c=2$ cm/s and dashed line to the flattened profile of $V\sb n$. }
 \end{figure} 

 To rationalize these observations we plot in  \Fig{f:6} the profiles of  various  velocities, normalized by the counterflow velocity $V^{\dag}=V/ V^0\sb {ns}$. We start with the streamwise component of the superfluid velocity $V^x\sb {s}(y)=V^0\sb s+V^x_{\sb{BS}}$. Near the walls    $V\sb s<0$ for all flow conditions and close to $V^0\sb s$. The main difference between the superfluid velocity behavior at low $T$ [panels (a) and (b)] and at high $T$ [panels (c) and (d)] is in the channel core, where at $T=1.3$~K $ V\sb s>0$, while at $T=1.9$~K $V\sb s<0$. As a result, at low $T$ the value of $V\sb{ns}$ is smaller than $U_c$, while at high  $T$, $V\sb {ns}$ is larger than $V\sb n$ everywhere in the channel and homogeneous across the core even for the parabolic $V\sb n$ profile. Furthermore, as is shown in \Fig{f:7}a, the shape of $V\sb{ns}(y)$ remain almost unchanged with increasing $H$ at low temperature while becoming flat over an increasingly larger part of the channel as the channel become wider at high $T$, \Fig{f:7}b. Since the tangle dynamics is defined by $V\sb{ns}$ according to \Eq{SFVel}, such  behavior may explain the tendency for a more homogeneous VLD distribution in wider channels at high $T$ than at low temperature.
  \begin{figure*}[htp]
  		\includegraphics[scale=0.43]{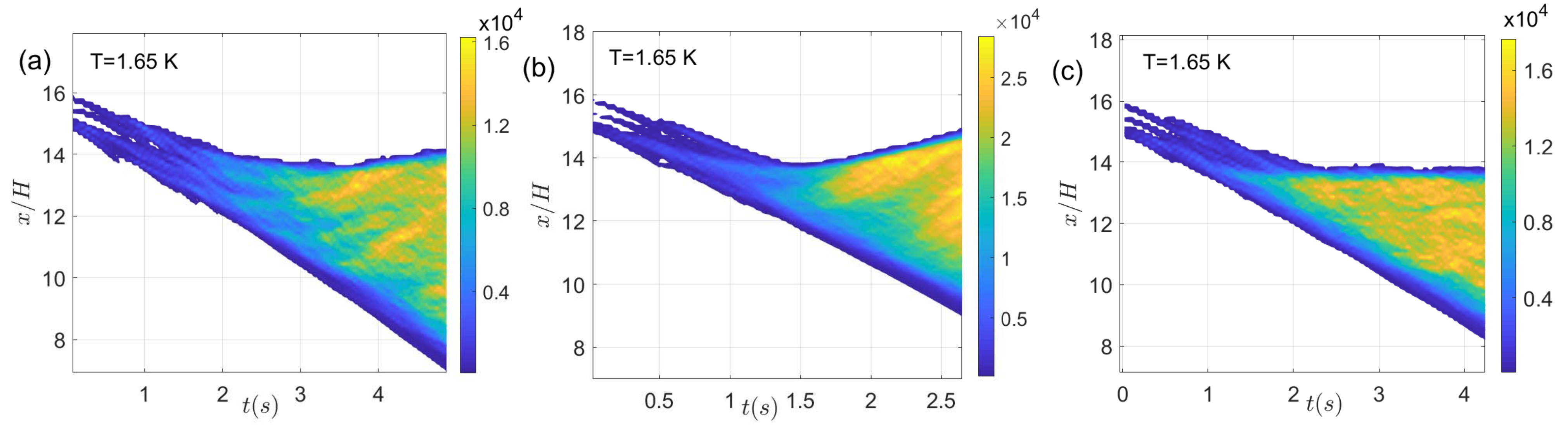}
	\caption{\label{f:9}VLD dynamics for $T=1.65$~K and various $V\sb n$ profiles: (a) parabolic profile wih $U_c=1.5$ cm/s, (b) parabolic profile with $U_c=2$ cm/s and (c) flattened $V\sb n$ profile, corresponding to (a).}
\end{figure*} 

 The superfluid velocity plays an additional role in the dynamics. It is usually assumed that the overall tangle motion is defined by the superfluid velocity $\B V^0\sb s$. However, as is shown in \Fig{f:3}b,d, the fronts of the tangle may both move in the direction of $\B V^0\sb s$, or only cold front moves with $\B V^0\sb s$, while the hot front moves in the opposite direction. It is natural  to associate the direction of the cold front motion with the direction of the  superfluid velocity $\B V\sb s$  near the walls, while the direction of the hot front motion with the direction of $\B V\sb s$ in the core of the channel. Such an assumption is further supported in \Fig{f:8}, where we plot the $V\sb s$ and $V\sb{ns}$ velocities for the intermediate $T=1.65$~K, and \Fig{f:9}, where we plot the evolution of the corresponding $\C L(x,t)$.  Here the superfluid velocity in the channel core is close to zero and the behavior of the hot front is very sensitive to the flow conditions. The superfluid velocity for the flattened $V\sb n$ profile, shown in \Fig{f:8} as the blue dashed line, is negligible at the center of the channel and the corresponding hot VLD front [\Fig{f:9}c] is stationary. The hot VLD front in the flow generated by the parabolic $V\sb n(y)$ with $U_c=1.5$ cm/s, for which $V\sb s(0)\lesssim 0$, has hardly settled [\Fig{f:9}a], despite relatively long propagation time. On the other hand, at larger $U_c=2$ cm/s, we clearly see in \Fig{f:8}b the hot front moving opposite to the direction of $\B V^0\sb s$.   The cold fronts under all conditions move with $\B V^0\sb s$, although the front speeds differ. Here, the cold front speed for the flattened $V\sb n$ profile is smaller than for the corresponding parabolic $V\sb n(y)$, consistently with a smaller value of $V\sb s$ at the walls. The fronts speeds are \emph{not equal} to $V\sb s$ at the wall or in the core, although they are clearly related.
 \begin{figure*}[htp]
 	 	\includegraphics[scale=0.44]{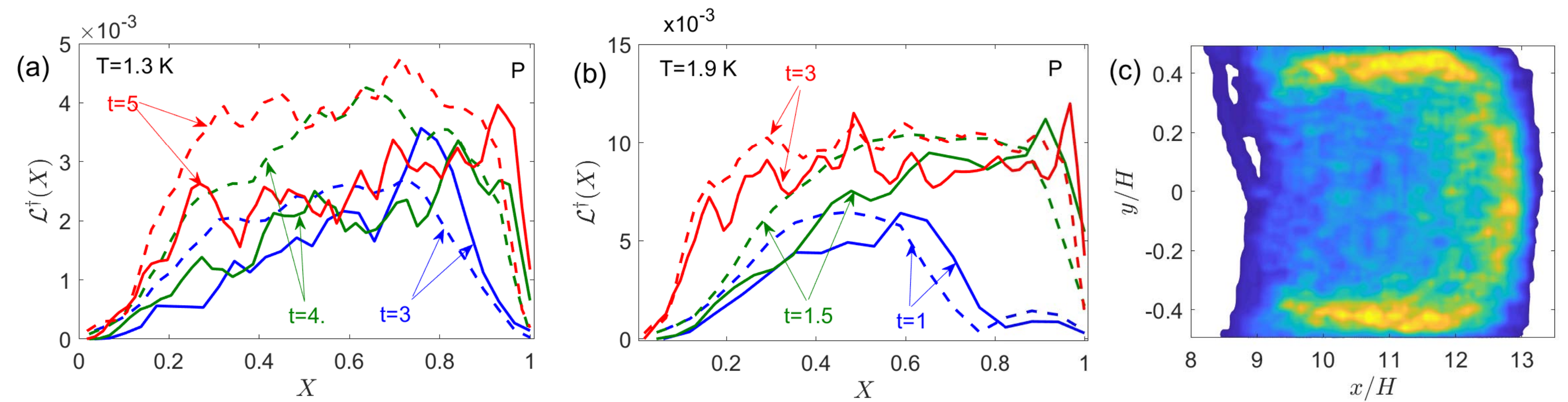}
 	\caption{\label{f:10}Rescaled normalized VLD profiles $\C L^{\dag}(x)$ for the parabolic $V\sb n$ at various time moments for (a) $T=1.3$~K, $U_c=3$cm/s and (b) $T=1.9$~K, $U_c=1$cm/s in the narrow channel $H=0.1$ cm. Solid lines denote the core VLD profile, dashed lines denote the wall profiles. (c) 2D VLD map $\C L(x,y)$ for $T=1.3$~K, $U_c=3$ cm/s, $H=0.15$ cm.   } 
 \end{figure*} 

 \begin{figure*}[htp]
 	\includegraphics[scale=0.45]{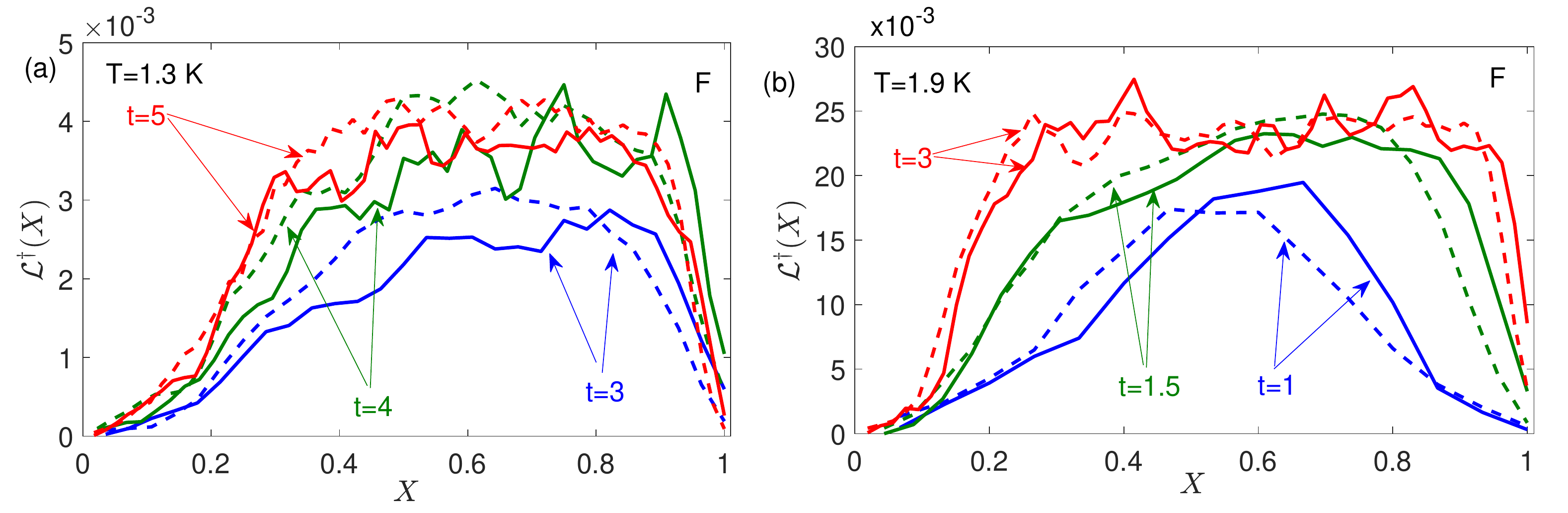}	
 	\caption{\label{f:11}Rescaled normalized VLD profiles $\C L^{\dag}(x)$ for the flattened $V\sb n$ at various time moments for (a) $T=1.3$~K,  $U_c=3$cm/s  and (b) $T=1.9$~K,$U_c=1$cm/s. Solid lines denote the core VLD profile, dashed lines denote the wall profiles.  } 
 \end{figure*}


   \subsection{\label{ss:trans}Transient Dynamics}
   In this Section, we consider the transient dynamics of the growing turbulent plugs for different conditions. Here we compare the changes in the shape of the tangle, plotting in \Figs{f:10} and \Fig{f:11} the dimensionless VLD $\C L^{\dag}(x)$ for the core and for the near-walls regions, rescaled to the tangle width at each of the presented three time moments. In this way, the scaled coordinate $X=0$ corresponds to the cold edge of the tangle and  $X=1$ corresponds to the hot edge. The earliest time moment corresponds to the time when the three-dimension (3D) tangle was formed and the latest to the time when the bulk region and two fronts are fully developed.

  The tangle dynamics for the parabolic $V\sb n$ profile is shown in \Fig{f:10}.
  The main feature of these profiles is the asymmetry  with respect to the center of the tangle.
  The wall profiles, shown by dashed lines, rise along all tangle length and the asymmetry is relatively mild. The core profiles, shown by solid lines,  on the other hand, are very asymmetric, with the hot side growing faster than the cold side. We can see that at $T=1.3$~K (\Fig{f:10}a) the growth of the cold side in the core is stalled compared to the walls profiles. This results in the wall-normal profiles  with a significant difference between VLD at the core and near the walls (cf. \Fig{f:5}a). Moreover, during all evolution, the core region leads in the hot front, while the wall region develops faster at the cold front. 
  
 Similar tendencies in the dynamics are observed at $T=1.9$\,K, \Fig{f:10}b. The main difference from the lower temperature regime is faster tangle development and closer values of $\C L^{\dag}$ in the wall and  in the core region, in accordance with \Fig{f:5}c. Notably, also here the core region first develops closer to the hot front (i.e in the direction of $\B V\sb n$), despite the fact that $\B V\sb s$ in the core is oriented  in  this case in the opposite direction.
 
 The main reason for this asymmetry is the spatial distribution of the driving velocity.  As is shown in Appendix \ref{a:3}, due  to enhanced   VLD production in the channel core in the hot front region, and the transverse VLD flux that moves  the vortex lines toward the walls, the parabolic wall-normal profile of the normal-fluid is translated into a transient VLD distribution that reminds a horseshoe shape: $\C L$ is higher near the walls and near the hot edge in the core of the channel, as is shown in \Fig{f:10}c. The clearly visible hump in the earliest  core VLD profile (e.g. blue solid line, labeled ``$t=3$'' in \Fig{f:10}a corresponds to the central part of the horseshoe. With the development of the tangle, the hump is redistributed  to the rest of the core region  and becomes less prominent, although it does not disappear  completely even when the bulk value of $\C L$ is established over a large part of the core. This horseshoe shape of the most dense part of the growing tangle lasts  longer for larger $U_c$ and wider channels. Such an asymmetry  of the tangle, that appears from the very beginning of the tangle development leads to very different initial conditions for the formation of plug fronts (see also  Appendix \ref{a:3}).
 
One may argue that such a scenario may not be realized in the  real  counterflow due to flattening of the normal-fluid profile and therefore more even initial VLD distribution. However, as we show in \Fig{f:11} and \Fig{f:prodX}(c,f), the streamwise tangle asymmetry is initially present even if the flattened $V\sb n(y)$ is imposed, with VLD growing faster at the hot edge of the plug. This transient behavior  does not last long in this case, however, the hot  front remains stepper than the cold front, similar to the tangles formed under  the parabolic  $V\sb n$.
    
        \begin{figure*}[htp]
    	\includegraphics[scale=0.41]{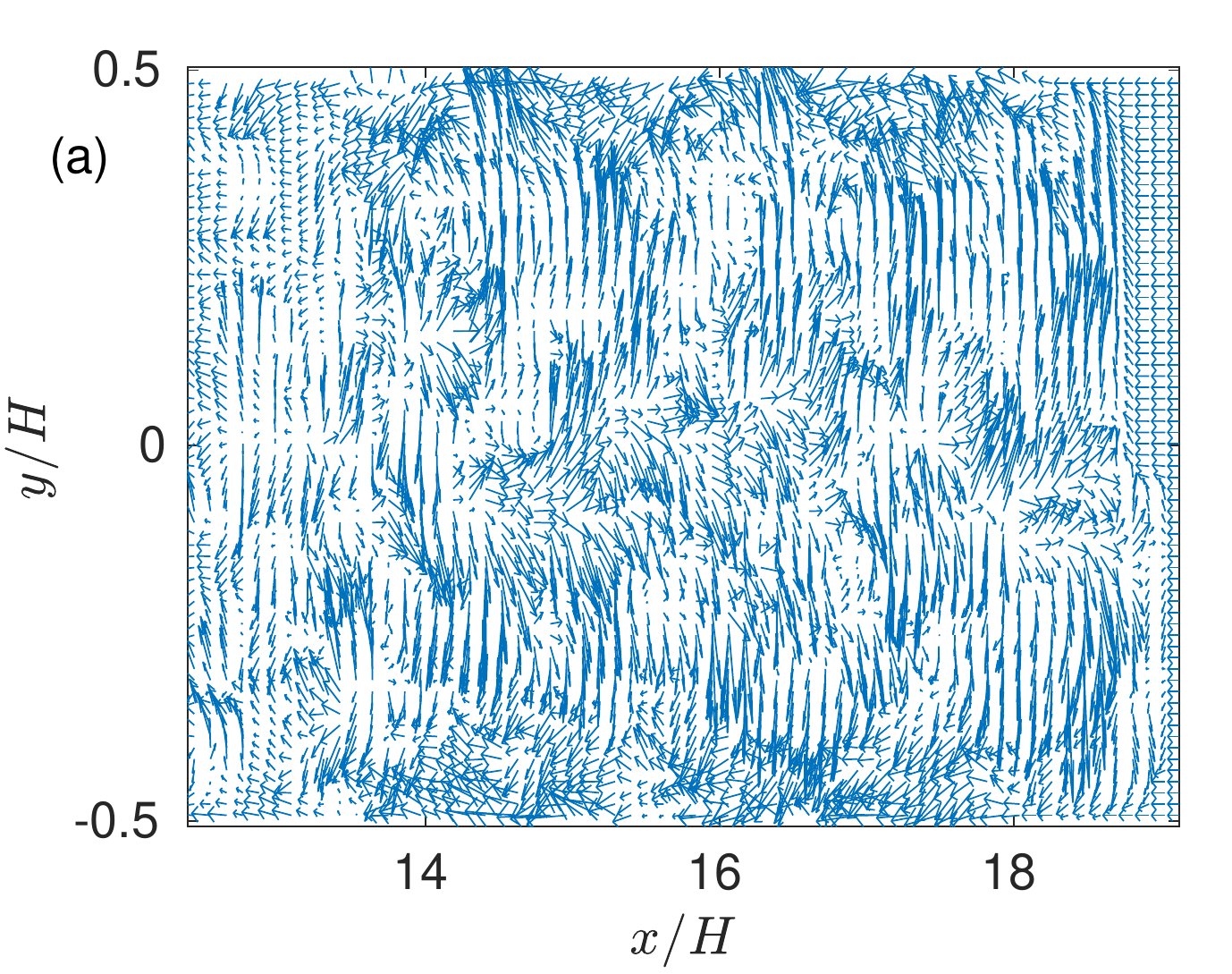}
    	\includegraphics[scale=0.41]{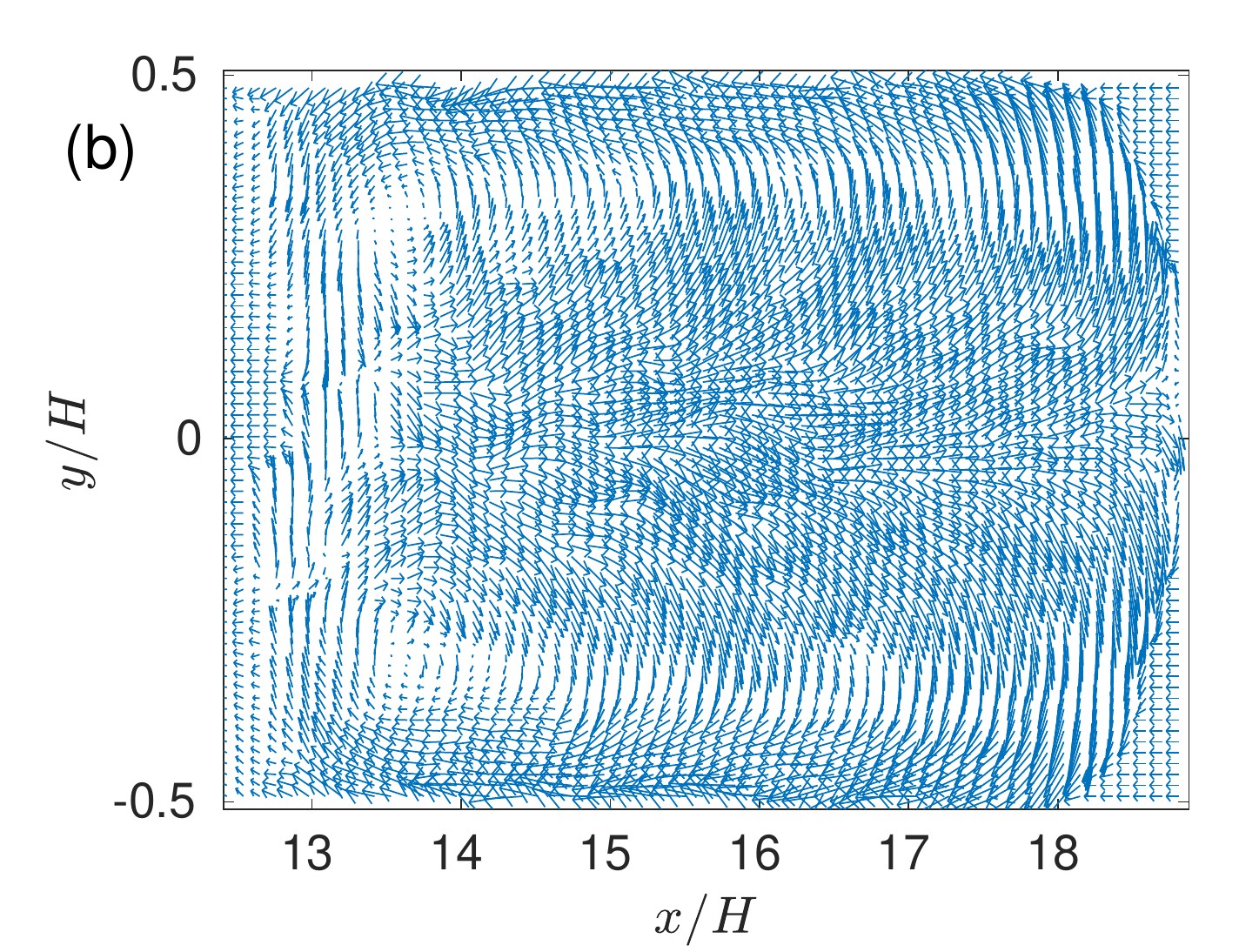}
    	\includegraphics[scale=0.41]{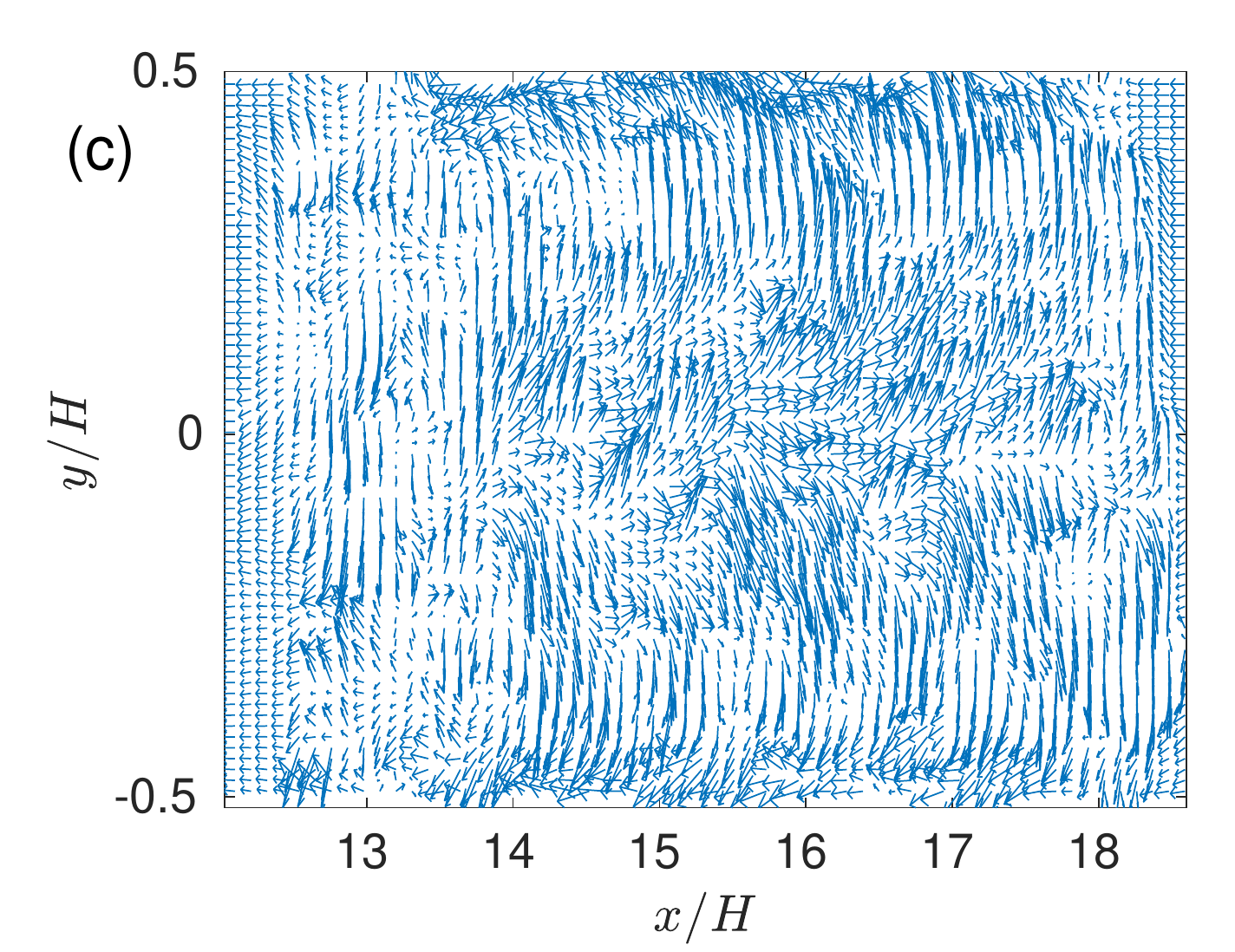}
    	\caption{\label{f:Vvect}Tangle drift velocity $\B V\sb{drfit}$ for $T=1.3$ K and (a) $U_c=2$ cm/s, (b) $U_c=4$cm/s and (c) flattened $V\sb n$ profile. The arrows direction shows the local orientation of the velocity, the size of the arrows is proportional to its magnitude.}
    \end{figure*} 

    \subsection{\label{ss:LargeScaleV}Large-scale superfluid motion}
    
    In simulations of homogeneous tangles under triply-periodic boundary conditions, the presence of mean normal-fluid and superfluid velocities is accounted for by a constant and space-homogeneous counterflow velocity, while the tangle-induced velocity is artificially randomized by interactions with image vortex lines.  In simulations of superfluid turbulence in the channel with periodic streamwise conditions, the translation invariance is broken in the wall-normal direction, creating superfluid motion from the center of the channel towards the walls. Still, in the streamwise direction, the variation of the vortex lines velocity is not taken into account.
    
    In our simulations, the tangle has finite streamwise length and the superfluid velocity varies along the tangle as well as across it. The drift velocity of vortex lines $\B V\sb{drift}$ that include the mean velocity as well as  all contributions of the tangle-induced velocity represent the superfluid motion at all scales that are formed in our system.  In \Fig{f:Vvect} we plot the tangle drift velocity for $T=1.3$~K, at which the mean superfluid velocity does not dominate and motion at all scales are clearly seen. Since near the wall the superfluid flows toward smaller $x$, while in the core its motion is oriented toward larger $ x$ values, eddies of various sides are formed. For parabolic $V\sb n$ profile, at weak driving velocity $U_c=2$ cm/s, \Fig{f:Vvect}a,  many circular eddies with sizes that are much larger than the intervortex distance $\ell$ but smaller than the channel size $H$, are formed.  At strong driving velocity $U_c=4$ cm/s, \Fig{f:Vvect}b,   two dominant vorticies  of the system size $H/2$  and opposite circulation orientation, covering  whole tangle length  are formed, with smaller motions masked by the largest ones. When the flow is driven by flattened $V\sb n$ profile, \Fig{f:Vvect}c, we can see both the system size motion and smaller eddies. Please note, that near the walls the tangle velocity contributions are oriented perpendicular to the channel walls due to no-slip boundary conditions. It is the mean superfluid velocity that moves the vortex lines near the walls and helps to create the large-scale eddies. At higher temperatures, the dominant  $\B V^0\sb s$ sweeps the tangle along the channel and masks the presence of smaller superfluid motions, similar to the sweeping velocity in classical fluids. However, analysis of the relative drift velocity $\B V\sb{drft}-\B V^0\sb s$ clearly shows the presence of these motions also in the vortex tangles at higher $T$.
  \begin{figure*}[htp]
	\includegraphics[scale=0.63]{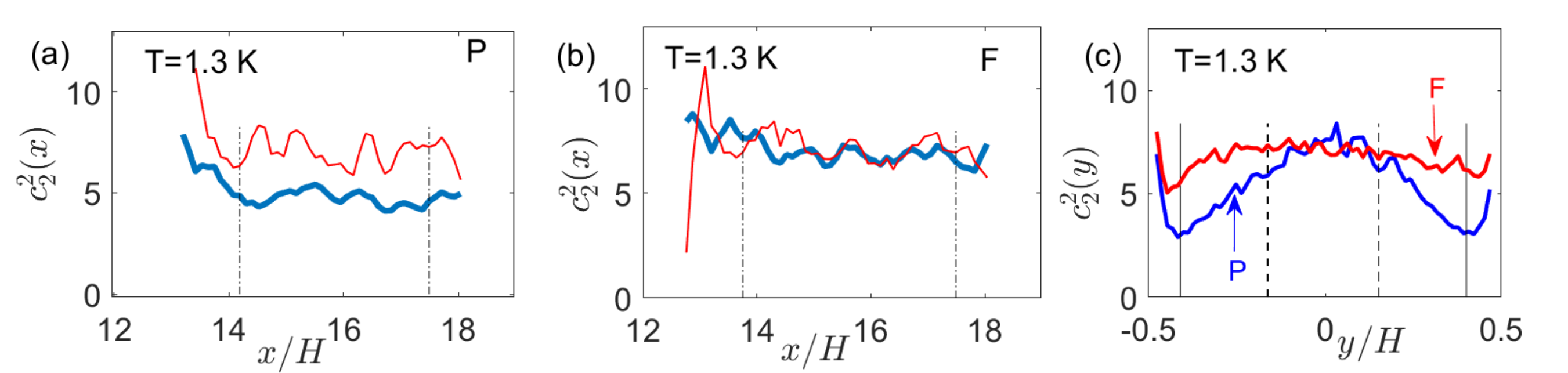}\\
	\includegraphics[scale=0.63]{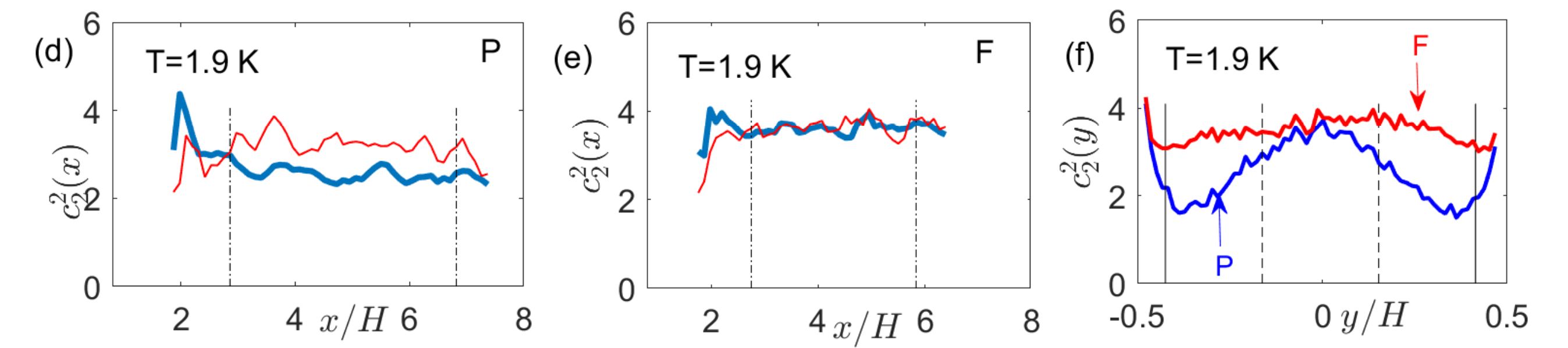}
	\caption{\label{f:c22} The coefficient $c^2_2$ at various conditions.  The profiles for $T=1.3$~K are shown in the top row: the streamwise profiles for (a)the parabolic $V\sb n$ with $U_c=3$ cm/s, labeled  ``P", (b) the   flattened profile, labeled ``F", (c) the corresponding wall-normal profiles.  The profiles for $T=1.9$~K are shown in the bottom row: (d) the parabolic $V\sb n$ with $U_c=1$ cm/s and (e) the flattened profile; (f) the corresponding wall-normal profiles.  Dot-dashed  and dashed black lines  mark the edges of the bulk and the core regions for the streamwise and for the wall-normal profiles, respectively. Thin solid lines in panels (c) and (f) are placed at the intervortex distance from the corresponding walls.  } 
\end{figure*} 

 \begin{figure*}[htp]
	\includegraphics[scale=0.62]{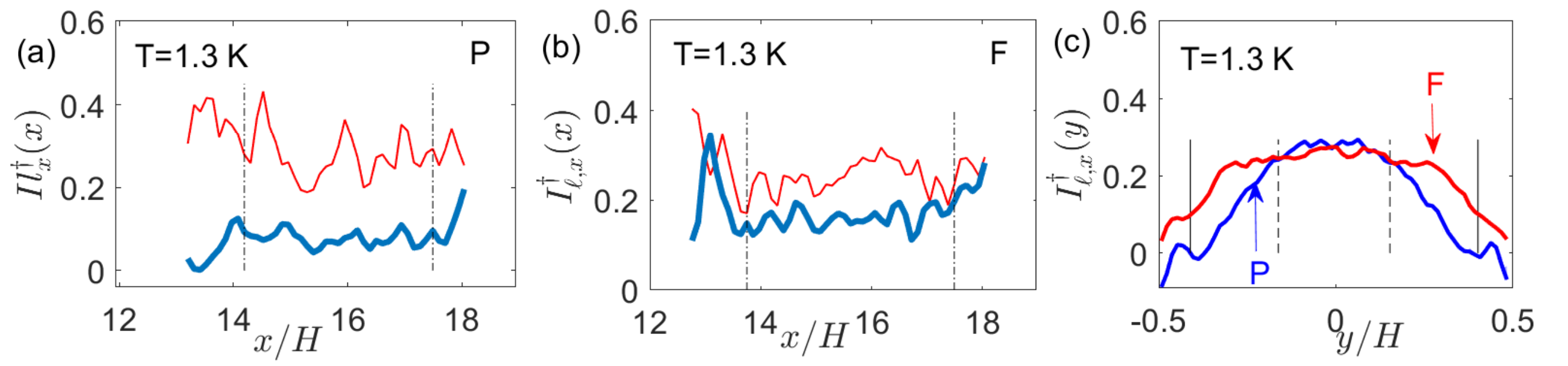}\\
	\includegraphics[scale=0.62]{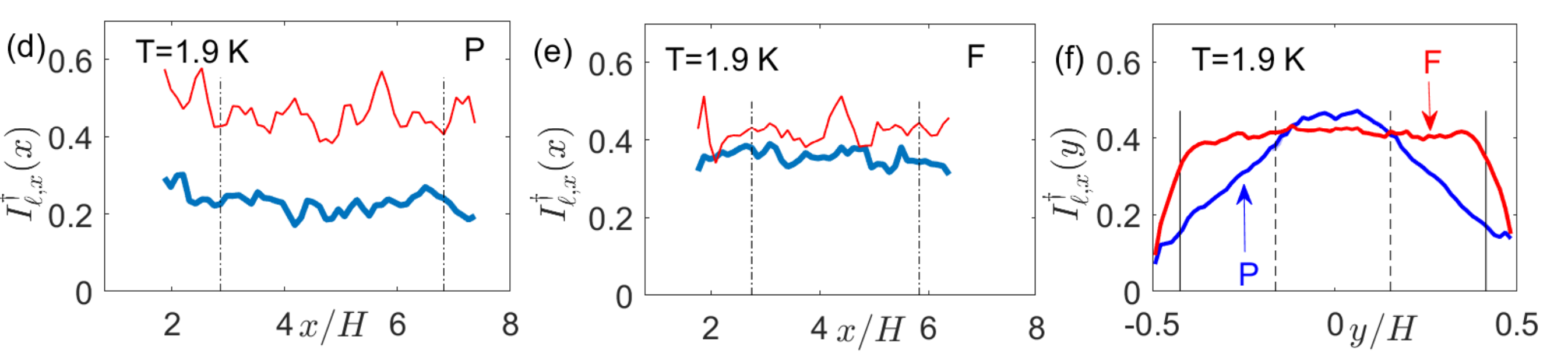}
	\caption{\label{f:ilx} The streamwise component of the  index  $I^{\dagger}_{\ell,x}$ at various conditions.  The profiles for $T=1.3$~K are shown in the top row: the streamwise profiles for (a) the parabolic $V\sb n$ with $U_c=3$ cm/s, labeled  ``P", (b) the  flattened profile, labeled ``F", (c) the corresponding wall-normal profiles.  The profiles for $T=1.9$~K are shown in the bottom row: (d) the parabolic $V\sb n$ with $U_c=1$ cm/s and (e) the flattened profile; (f) the corresponding wall-normal profiles.  Vertical dot-dashed lines  mark the edges of the bulk of the streamwise profiles,  vertical dashed lines mark the edges of the core  for the wall-normal profiles. Thin solid lines in panels (c) and (f) are placed at the intervortex distance from the corresponding walls.  } 
\end{figure*} 

   \begin{figure*}[htp]
   \includegraphics[scale=0.6]{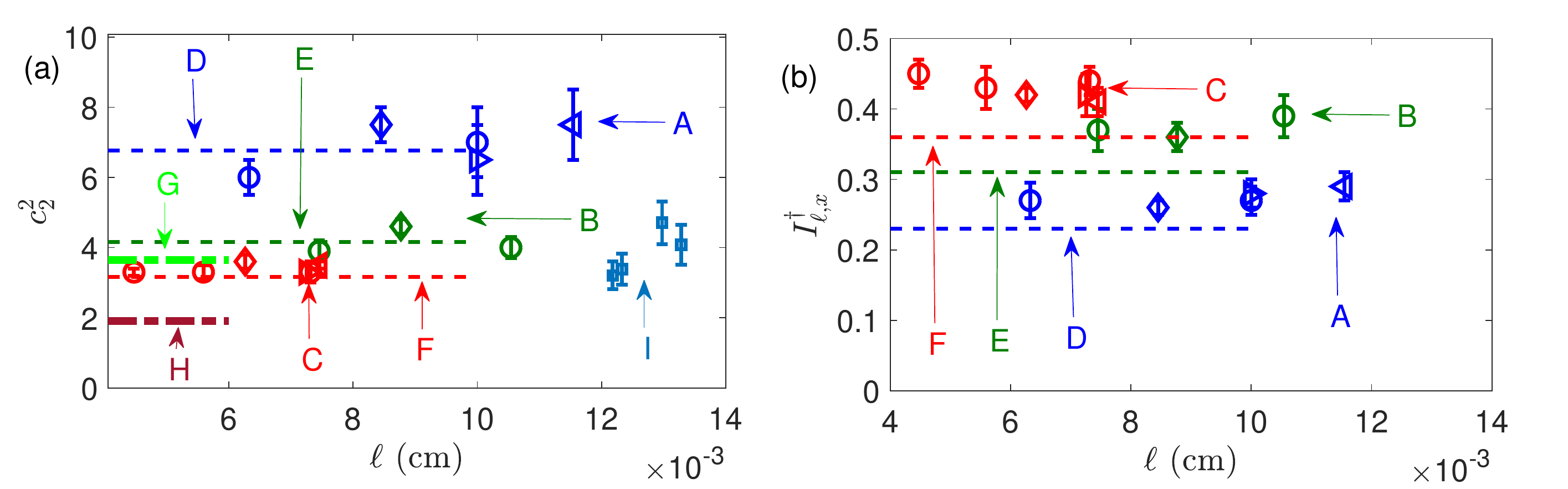}
	\caption{\label{f:c22all} The mean values of (a) $c^2_2$ and  (b) $I^{\dagger}_{\ell,x}$ in the core of the channel. The labels (A-C) correspond to current simulations (A: $T=1.3$ K,blue symbols, B: $T=1.65$ K, green symbols, C: $T=1.9$K, red symbols). Different flow conditions are represented by different symbols: $\circ$ denote front velocities for the  parabolic $V\sb n$ and various $U_c$, $\diamond$ corresponds to the flattened $V\sb n$ profile, $\triangleright$  and $\triangleleft$ denote channel width $H=0.15$ cm and  $H=0.2$ cm, respectively.  The error-bars denote  standard deviation from the mean in the bulk of the tangle.  Dashed horizontal lines, labeled D-F ($T=1.3, 1.6$ and $1.9$ K, respectively) are the values of $c^2_2$ and $I^{\dagger}_{\ell}$ for the homogeneous tangle from \Ref{recon14} with the GEC reconnection criterion, the same as used in this paper.  In panel (a), thick dot-dashed lines, labeled G ($T=1.65$ K)  and  H ($T=1.95$K), denote the experimental values of  $c^2_2$ from  \Ref{SkrbekVarga18}  in the range of intervortex distances $[4\times 10^{-3}-6\times 10^{-3}] $cm. Filled squares with error-bars, labeled "I", denote the results of simulations in the channel with parallel solid plates from  \Ref{WeiTsuVinen18} in the range of temperatures $T=1.4-1.7$K.}
\end{figure*}

      \subsection{\label{ss:c22}Structural parameters $c^2_2$ and $\B  I^{\dagger}_{\ell}$}
     In the microscopic description\cite{schwarz88} of the tangle dynamics very important role is played by two structural parameters: the local binormal $\B I_{\ell}=\langle \B s' \times \B s''\rangle$ and the ratio between the vortex line density and the mean-square curvature $c^2_2$. These parameters contribute to the terms of the equation of motion for $\C L$, responsible for the production and annihilation of the vortex line length, respectively [cf. \Eqs{terms}-\eqref{decay}].   In the homogeneous tangles, these parameters are constants, while in the channel flow they depend on the position in the channel. The behavior of these parameters at the edges of the tangle was not studied so far.
     
    The profiles of the coefficient $c^2_2$ are shown in \Fig{f:c22} for $T=1.3$~K [panels (a)-(c)] and for $T=1.9$~K [panels (e)-(f)].  There are several common properties of the streamwise profiles [panels(a),(b),(d) and (e)], independent of the temperature and the type of the driving velocity. The values of $c^2_2$  exhibit fluctuations along the tangle with a relatively large amplitude, especially when the flow is driven by the parabolic $V\sb n$. The fluctuations are less pronounced in the wall-normal profiles [panels(c) and (f)]. These profiles have a somewhat different averaging scope, however, we incline to attribute these fluctuations to  the streamwise inhomogeneity  of both VLD and the curvature, that do not match exactly. Nevertheless, the values of $c^2_2$ are fairly constant along the tangle, with the same values observed also the hot front region. The behavior of  $c^2_2$ in the cold front regions is different, with the tendency of becoming larger at low $T$. Similar behavior is observed for other values of $U_c$ (not shown). 
    
     When the flow is driven by the flattened $V\sb n$ profile, the values of $c^2_2$ may be considered almost constant  across the channel. Its behavior changes only within the intervortex distance from the wall, where the values of VLD drop very quickly, while the square curvature keeps its values almost until the wall. Conversely, when the driving velocity has parabolic profile, $c^2_2$ has the largest values in the center of the channel and decreases linearly towards the walls until the  intervortex distance is reached. Then it increases, in a similar way as for the flattened $V\sb n$ profile, even reaching similar values at the wall. These larger values in the core of the channel, as compared to the near-walls region, are observed along all the tangle bulk and in the hot front. On the other hand, the values of $c^2_2$ in the flows driven by flattened $V\sb n$ may be considered almost space-homogeneous, except for the cold front and very near the walls, and similar to the values of $c^2_2$ in the channel core, observed for the parabolic normal flow.
     
     The dominant contribution to the production of vortex lines has the term\cite{schwarz88} that depend of the streamwise projection of the local binormal $I_{\ell,x}$. In \Fig{f:ilx} we plot its values normalized by the  mean curvature  $I^{\dagger}_{\ell,x}=I_{\ell,x}/\varkappa$. The general behavior of $I^{\dagger}_{\ell,x}$ is similar to that of $c^2_2$. We therefore point out main differences. 
  Looking at the wall-normal profiles, \Fig{f:ilx}c,f, we notice that $I^{\dagger}_{\ell,x}$ is almost homogeneous over the channel core for parabolic flows at both temperatures, crossing over to a linear decrease toward the walls beyond the core region. It does not increase significantly very near the wall, although at $T=1.3$\,K a kink is observed. This kink becoms stronger for wider channels and appears at $T=1.9$\,K for wide channels as well. For the flow, driven by the flattened normal-fluid velocity profile, the core values of $I^{\dagger}_{\ell,x}$ extend further toward to walls, especially at high $T$. Nevertheless, the difference between the mean values of the channel core and the near-walls region persists along the tangle bulk, even in this case. The values of  $I^{\dagger}_{\ell,x}$  in both fronts regions differ from the bulk, even if we take into account  strong fluctuations in its streamwise distribution.
  
  Interestingly, the shape of wall-normal profiles of $c^2_2$ and $I_{\ell,x}$  in the flows, generated by the flattened $V\sb n$ profiles, does not depend on $V\sb{ns}$ and the channel width at both high and low $T$, although for different reasons.   At low $T$, the curvature is only weakly dependent on the distance from the wall, while  VLD strongly peaks near the walls. At high $T$, wall-normal profiles of $\C L$ are more homogeneous, but the curvature, in this case, decreases toward the walls more strongly. The resulting $y$-distributions of $c^2_2$, \Fig{f:c22}c,f, are very similar.  The wall-normal distribution of  $I^{\dagger}_{\ell,x}$  is fully defined by the streamwise component of the binormal that is large in the center of the channel and quickly decreases toward the walls. Its shape is only slightly altered by similar distribution of $\langle |s^{\prime \prime}|\rangle $, \Fig{f:ilx}c,f.
  
  To get an idea of how these results are related to other known measurements, we compare in
     \Fig{f:c22}a the values of $c^2_2$ for the channel core with the results of simulations of the homogeneous  tangles\cite{recon14}, in the planar channel\cite{WeiTsuVinen18}  and with the experimental results \cite{SkrbekVarga18}  for  the range of intervortex distances, typical for our simulations.   We have chosen to compare the values for the core of the channel because the experiments were carried out in wide channels, where the core behavior is expected to dominate. These values are also expected to be more comparable with $c^2_2$  in the homogeneous tangle.
    
    As is clearly seen, the calculated values of $c^2_2$ do not depend on the intervortex distance within the range used in our simulations. The temperature dependence agrees with previous results, i.e. larger $c^2_2$ at lower temperatures. Our current results agree well with the values obtained in the homogeneous vortex tangles\cite{recon14}, shown by thin dashed lines.  The values of $c^2_2$ obtained in numerical simulations of the vortex tangle  in the flow between parallel plates\cite{WeiTsuVinen18} for temperatures between $1.4-1.7$ K are shown by filled squares. The values of $c_2$ were calculated as averages over the whole channel and are expected to be lower than the values in the channel core. With this in mind, they agree reasonably well with our results for $T=1.65$K.
    
    The experimental values calculated using the fit\cite{SkrbekVarga18} for $T=1.65$K, shown by thick dot-dashed lines, are somewhat smaller, but not far from the numerical results. Here we need to take into account that  in the considered range $\ell=4\times 10^{-3}-6\times 10^{-3}$ cm, the fit becomes unreliable and the experimental points tend to scatter, see Fig.11 in \Ref{SkrbekVarga18}. The experimental values for $T=1.95$K are expectedly lower than our results for $T=1.9$K.
    
    Similar measurements of  $I^{\dagger}_{\ell,x}$ in the channel core are shown \Fig{f:c22}b. Note that this parameter measures alignment of the local velocity $\B V\sb{loc}$ with the direction of the counterflow velocity. Similar to $c^2_2$, the index $I^{\dagger}_{\ell,x}$   is fairly constant for a given temperature,  being larger for higher $T$.  These values are somewhat higher than those obtained in the homogeneous tangles\cite{recon14}, including the values obtained in the flow driven by the flattened normal-fluid profile (marked by diamond symbols). The temperature mismatch ( $T=1.65$\,K in our simulation vs $T=1.6$\,K in \Ref{recon14}) may account for the difference at the intermediate temperature, however, the trend is systematic across the temperatures.

    \section{\label{s:fronts}Front dynamics and analysis of the VLD balance equation}  
    \subsection{\label{ss:backgr}Background Overview}
Interface motion and front propagation in fluids are subject of intensive studies in various fields of knowledge.
Perhaps most well known are chemical
reaction fronts in liquids\cite{KPP37}, population dynamics of ecological communities\cite{Fischer37}, and combustion\cite{Zeldovich38}.  
 The mathematical description of those phenomena is
 based on partial differential equations (PDE) for the evolution
 of the concentration of the reacting species and the evolution
 of the velocity field. The two PDEs for the reactants and the velocity field are usually coupled, often in a nontrivial way. A mathematical simplification can be obtained by  neglecting the back-reaction of the reactant on the velocity field, which evolves independently. Such simplification is usually justified for the  laminar velocity field.  Even in such a limit, the front  dynamics is still  nontrivial and it is described by
 a so-called \emph{advection-reaction-diffusion} (ARD) equation
\begin{equation}\label{ARDE}
\partial \theta /\partial t + \B u(\B r,t)\cdot \B \nabla \theta= D \B\nabla^2 \theta + F(\theta),
\end{equation}
where $\theta(\B r, t) \in [0,1]$ is the reactant concentration, $D$ is the diffusivity and $F(\theta)$ is the reaction  term. 
The front interface is in general two-dimensional, although in many cases it is sufficient to consider its motion only in one direction. In a  typical model situation the localized initial conditions are used, i.e.  $\theta(\B r,0)\to 1$ exponentially fast when $\B r \to -\infty$ and $\theta\to 0$ exponentially fast when $\B r \to \infty$. In this case, the reaction front will move towards positive $r$. Here $\theta=0$ is an unstable state and $\theta=1$ a stable one, therefore $F(\theta)$  satisfies the condition 
\begin{equation}\label{FthetaCond}
F(0)=F(1)=0\, ,\quad F(\theta)>0 \, , \textrm{if} \quad 0<\theta<1\, .
\end{equation}
 It was shown\cite{vanSaarloos88,vanSaarloos2000} that if there is no advection, the front speed converges to a limiting velocity $v_0$, defined by a marginal stability condition. In a moving fluid, it is natural to expect \cite{Solution,Abel2001} that the front will propagate with an average (turbulent) speed $v\sb f>v_0$.
The turbulent front speed $v\sb f$ is defined by relative importance of the
flow characteristics, such as the relevant system size $\Lambda$, advecting velocity $u$, the diffusivity $D$, and the typical  time scale
$\tau_r$ of the reaction term $F(\theta)= f(
\theta)/\tau _r$.  The shape of $F(\theta)$, or more specifically the value $\theta$ at which it has largest slope, also plays a very important role. Two types of its functional dependence are of particular importance: 
(1) a Fischer-Kolmogorov-Petrovskii-Peskunov (FKPP) nonlinearity\cite{Fischer37,KPP37} $F(\theta)=\theta(1-\theta)$, or in general, any convex function $F''(\theta)<0$; (2) an Arrhenius (or ignition) nonlinearity\cite{Abel2001} $F(\theta)=\exp^{-\theta_c/\theta}(1-\theta)$. Here the parameter $\theta_c$ is an activation concentration, below which there is almost no production. 

In case of FKPP nonlinearity, the maximum slope $F(\theta)$ occurs at $\theta=0$. Such fronts are called \emph{pulled} fronts and their dynamics is fully determined by the region $\theta \approx 0$, as if pulled by the leading edge.  When the maximum  slope of $F(\theta)$ occurs at $\theta>0$, the front is \emph{pushed} by the non-linear interior.
 The allowed velocity of the pulled fronts has to satisfy the condition\cite{KPP37,Solution}
 \begin{equation}
 2\sqrt{D F'(0)}\le v\sb{min}< 2\sqrt{D \sup_{\theta}  \frac{F(\theta)}{\theta}}\, ,
 \end{equation}
  where $F(\theta)/\theta$ is the measure of the growth rate. For FKPP dynamics $F(\theta)/\theta=F'(0)$ and  for localized initial conditions $v\sb{min}=v_0=2\sqrt{D F'(0)}$. 
  
  For pushed fronts, the minimal front velocity $v\sb{min}$ is always larger than $v_0$. In both cases,  depending on the steepness of the initial conditions the asymptotic front speed may relax to the minimal $v\sb{min}$ or remain larger.
 
  There exists a vast literature on the front propagation in various flows. We concentrate on the laminar shear flow of ADR type and summarize several important results. For details see \Refs{Solution,Abel2001} and references therein.
  
  \begin{itemize}
  	
 	\item The front velocity is bounded by $K_1 u< v\sb f< v_0+K_2 u$, where the limiting velocity  $v_0=2\sqrt{D_0\sup_{\theta} [F(\theta)/\theta]}$   and  the diffusivity $D_0$ are the parameters in the absence of the advecting flow,  $K_1,K_2$ are flow-dependent parameters.
 		\item The diffusive transport is enhanced by the incompressible flow, resulting in an effective diffusion coefficient $D\sb{eff}>D_0$.
 		
 	\item In the presence of advection, the bound on limiting velocity may be modified as $v\sb f\le2\sqrt{D\sb{eff}  \sup_{\theta} [F(\theta)/\theta}]$.
 	
 	

  	
  \end{itemize}
   \begin{figure*}
  	\includegraphics[scale=0.52]{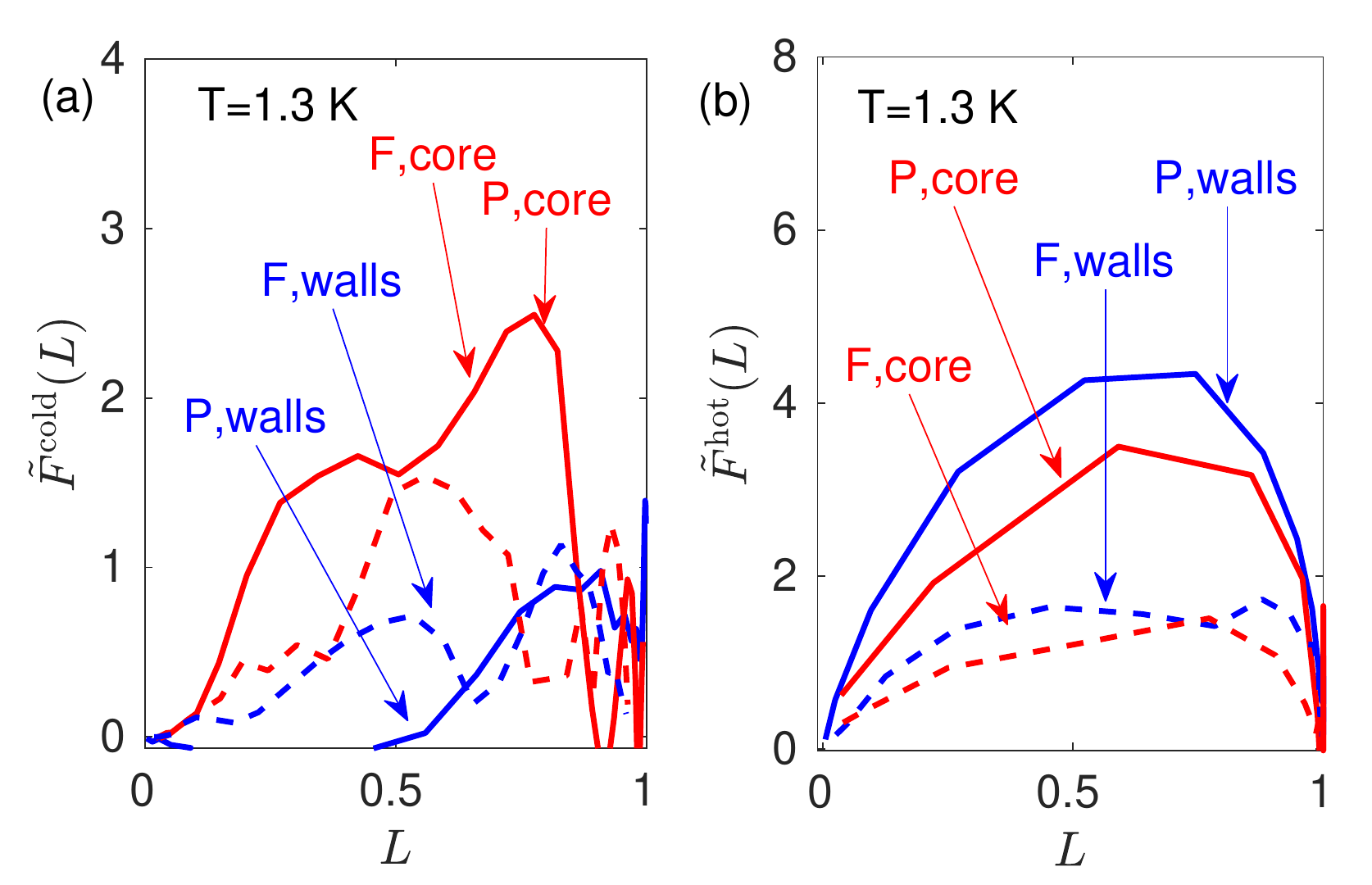}
  	\includegraphics[scale=0.52]{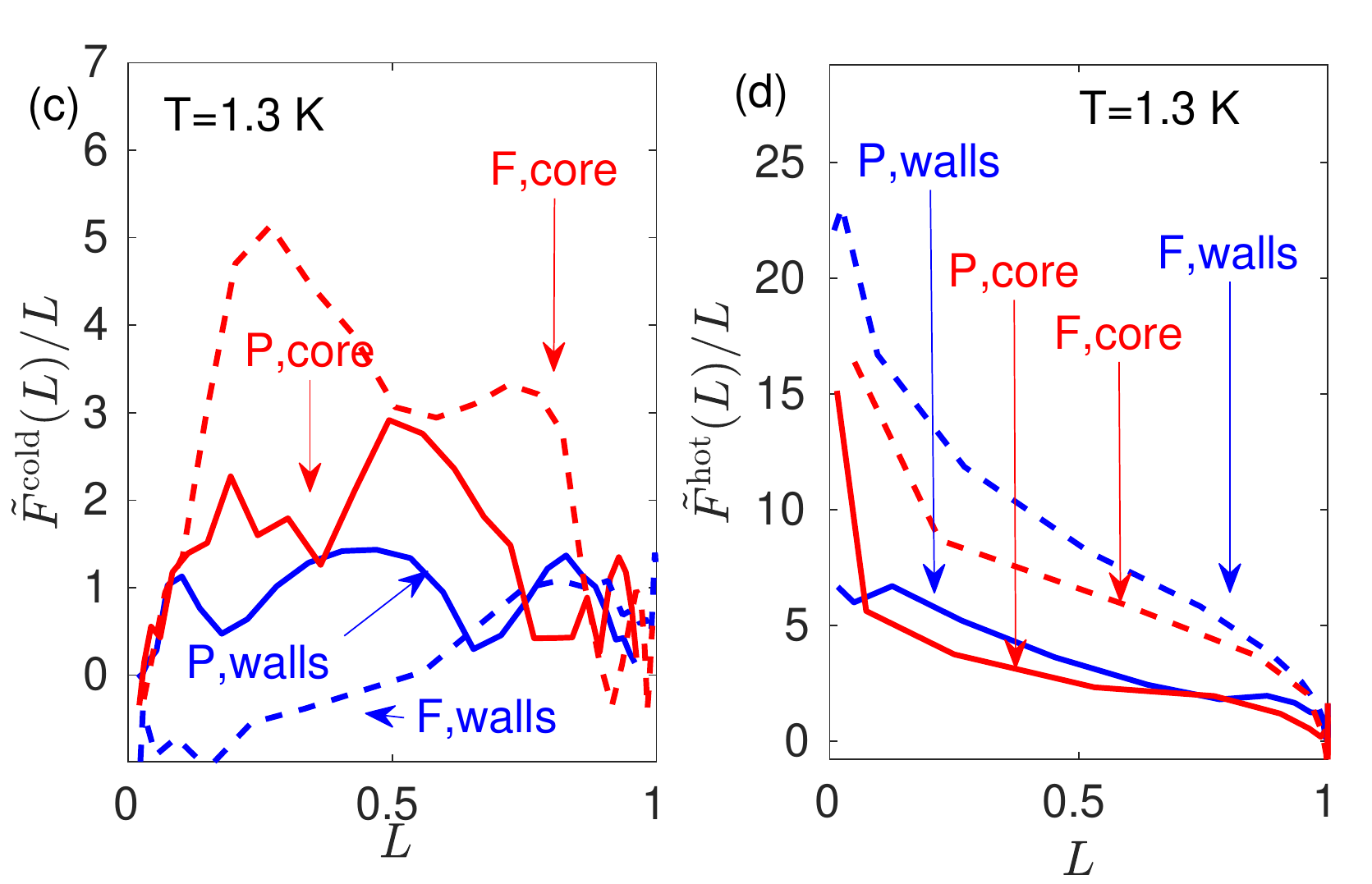}
  	\caption{\label{f:13}  The  profiles of $\tilde F^j(L)$ vs $L$  [panels (a) and (b) for the  cold and hot fronts, respectively] and $\tilde F^j(L)/L$  vs $L$ [panels (c) and (d) for the cold and hot fronts, respectively] for $T=1.3$~K. The profiles for the parabolic $V\sb n$ are shown by solid lines, the profiles for the flattened $V\sb n$ are shown by dashed lines and denoted as ``P" and ``F", respectively. The lines for the channel core and for the walls region are labeled in the figure.}
  \end{figure*}

   \begin{figure*}
  	\includegraphics[scale=0.39]{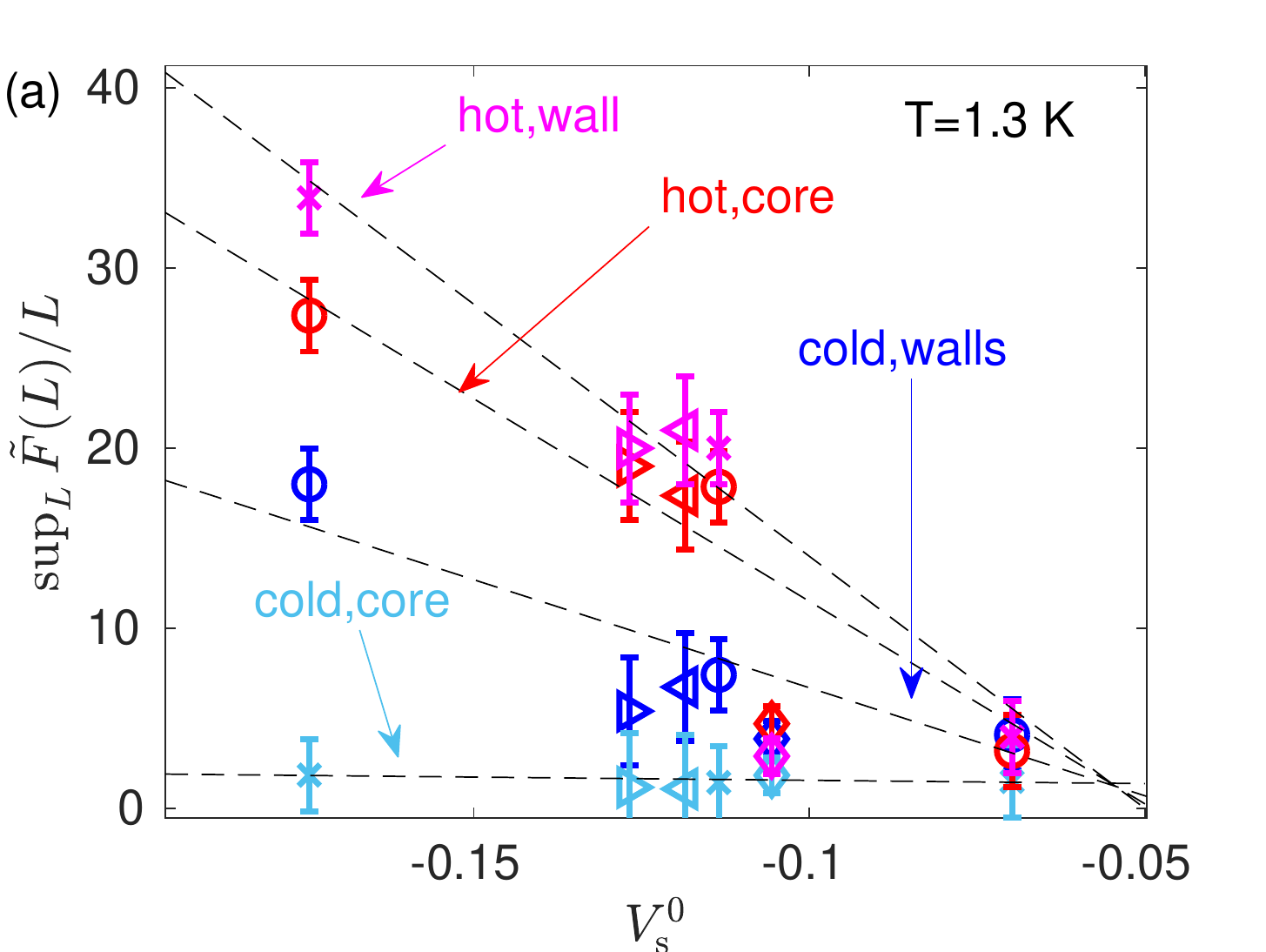}
  	\includegraphics[scale=0.39]{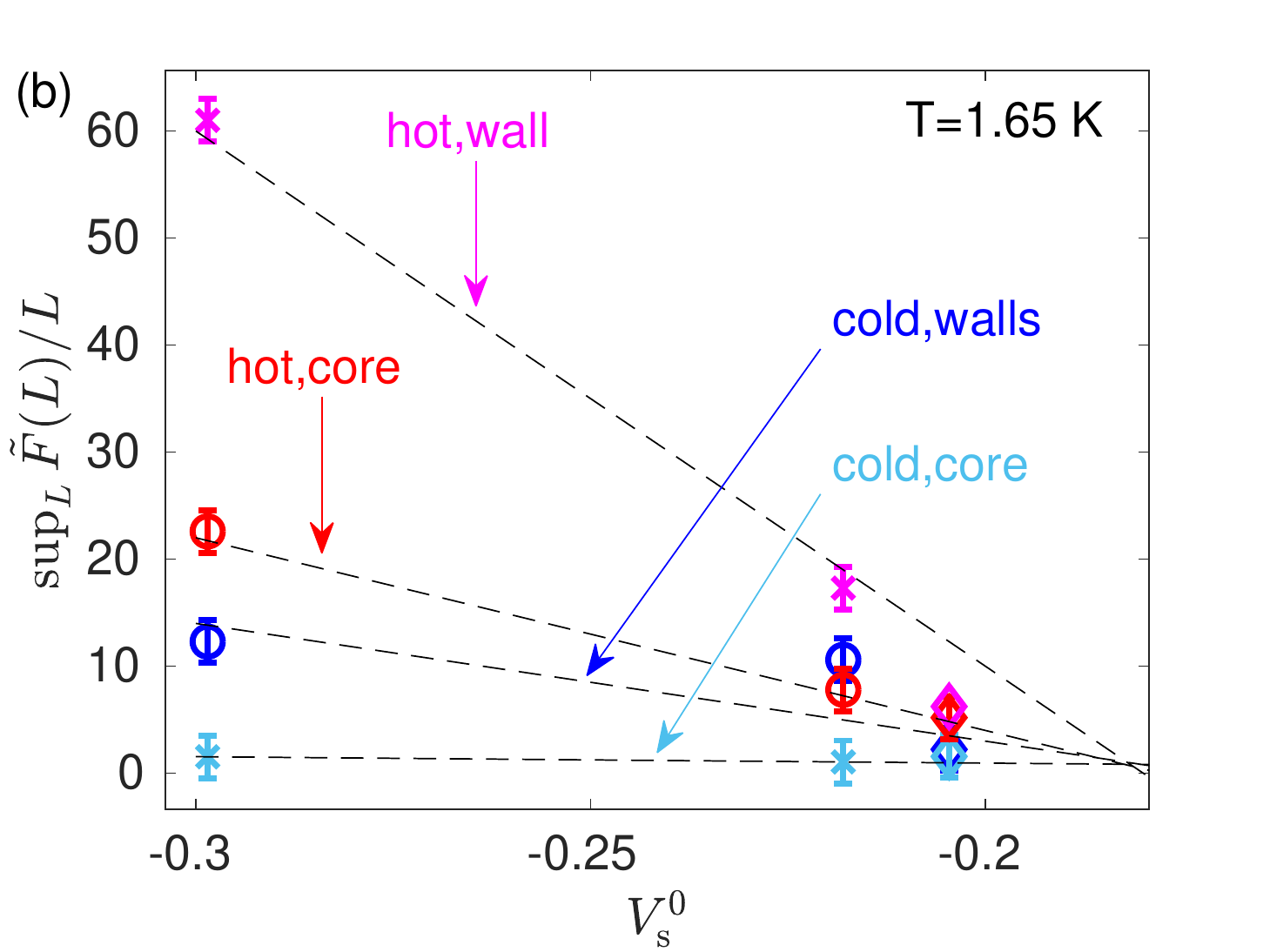}\
  	\includegraphics[scale=0.39]{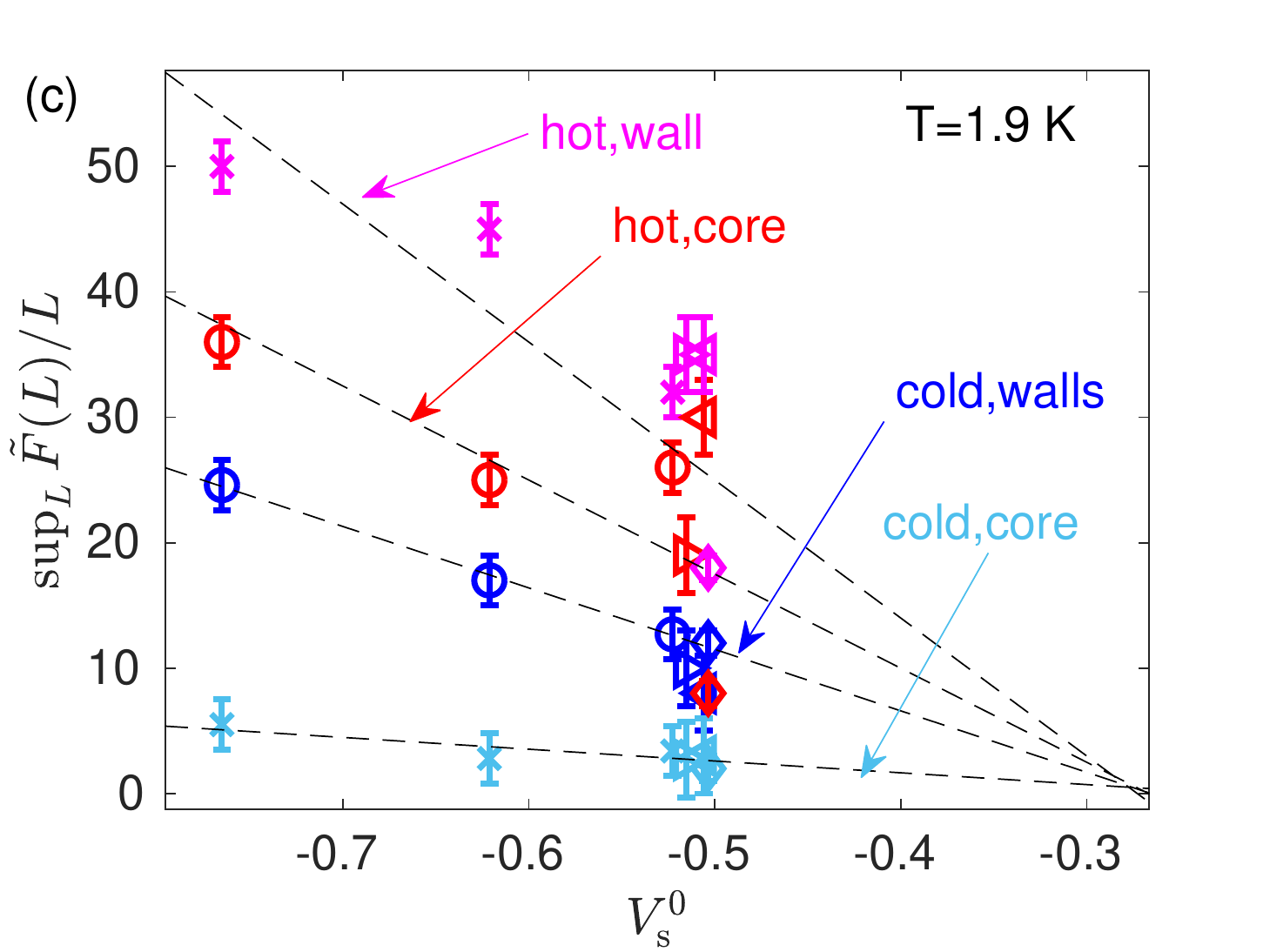}
  	\caption{\label{f:14} Maximum growth rate $\sup_L[ \tilde F^j(L)/L]$ at various flow conditions. In all panels, $\circ$ denote front velocities for parabolic $V\sb n$ and various $U_c$, $\diamond$ corresponds to the flattened $V\sb n$ profile, $\triangleright$  and $\triangleleft$ denote channel widths $H=0.15$ cm and  $H=0.2$ cm, respectively.  The linear dependence on $V^0\sb s$ is shown by dashed lines, which serve to guide the eye only. Different data sets are marked in the figure by labels of the same color that point to the corresponding symbols.   }
  \end{figure*}

  \subsection{\label{ARDEQ} ARD-type equation for VLD. }
  
  Now we return to the channel counterflow of the superfluid  $^4$He and relate the properties of the model system, described in the previous section, to the dynamics of the turbulent vortex tangle.
  
   Here the role of the dimensionless variable $\theta$ in advection-reaction-diffusion equation \eqref{ARDE}  is played by the normalized VLD $L=\C L/\C L_0$, where $\C L_0$ is the equilibrium vortex line density in the bulk of the tangle. The equation of motion for $L\B (r,t)$ in the channel may be written as
   	\begin{equation}\label{mainEq}
   \partial_t  L(\B r ,t) +\B \nabla[\B  V\sb {drift} (\B r) L(\B r ,t) ]= \tilde D \B \nabla^2   L(\B r ,t) +F[L( \B r,t)]\, ,
   \end{equation}
where $\tilde D$ is the effective diffusivity of VLD and $\B V\sb {drift}$ is the tangle drift velocity, see \Eq{SFVel}.
 We follow Schwarz's microscopic approach \cite{schwarz88} and recall  that the rate of elongation of the vortex line segment $\delta \xi$ is
   \begin{eqnarray}\label{segmentlength}
   \frac{1}{\delta\xi}\frac{d\delta\xi}{dt}&=&\alpha (\B V\sb{ns}(\B s, t)\cdot (\B s^{\prime} \times \B s^{\prime \prime})-  |\B s^{\prime} \times \B s^{\prime \prime}|^2  )\\ \nonumber
 & +&\B s^{\prime} \cdot{\B V\sb{nl}}^{\prime}-\alpha^{\prime}\B s^{\prime \prime}\cdot \B V\sb{ns}\, .
   \end{eqnarray}
  Integration of \Eq{segmentlength} over the vortex tangle gives  for the right-hand-side (RHS) term $F(L)$

  \begin{eqnarray}\label{terms}  
  F&=&\C P_1+\C P_2+\C P_3-\C D\, ,\\\label{p1}
  \C P_1 &= &\frac{\alpha}{\C L_0 V'}  \int_{\Omega'}  (\B V^0\sb{ns}-\B V\sb{nl}) \cdot (\bm s'\times \bm s'') ~d\xi  \,, \\\label{p2}
  \C P_2&=&\frac{1}{\C L_0  V'} \int_{\Omega'}\B  s^{\prime}\cdot  \B V\sb{nl}^{\prime }~d\xi  \, ,\\\label{p3}
  \C P_3&=&- \frac{\alpha'}{\C L_0  V'}  \int_{\Omega'} \B s^{\prime \prime}\cdot \B V\sb{ns}\, d\xi \, , \\  \label{decay}
  \C D &= &\frac{\alpha}{\C L_0  V'}   \int_{\Omega'}\B V\sb {loc} \cdot (\bm s'\times \bm s'') d\xi  \, .
  \label{EqL}
\end{eqnarray}
Here $\C P_1$ is usually named the production term since it is responsible for most of the vortex line elongation. The last term $\C D$ is traditionally termed the decay term since it represents the annihilation of vortex-line length during vortex dynamics and reconnections. Two other terms $\C P_2$ and $\C P_3$ also represent the production of the vortex-line length.
In the homogeneous, tangle $\C P_3$ vanish by symmetry. The term $\C P_2$ is usually omitted due to smallness. We include  all terms since $\C P_2$ and $\C P_3$ become non-negligible at low $T$ near the walls (see Appendix \ref{a:3}).  Each term is proportional  to $L$ due to integration over $d\xi$ and division by $\C L_0$. At this stage, we retain the integral representation of $F(L)$.

Using the same approach, the VLD flux is defined as
\begin{equation}
\C J=\frac{1}{\C L_0 V'} \int_{\Omega'}\B V\sb{drift}~d\xi= \B V^0\sb s L+ \frac{1}{\C L_0 V'} \int_{\Omega'}\B (V_{\sb{BS}}+\B V\sb{mf})~d\xi \, .
\end{equation}

As was shown in \Sec{ss:dyn}, the bulk VLD and other tangle properties in the core of the channel and near the walls are different but well defined. Therefore, instead of taking into account  full 3D structure of the tangle, as well as 2D front interface, we consider the dynamics of the core and the wall regions separately as one-dimensional (1D).

However, to get 1D equation for $L(x)$, it is not sufficient to only account for the  streamwise component of \Eq{mainEq}. Although the transverse diffusion is negligible, the transverse VLD flux $\C J_y $ is an important factor in the inhomogeneous tangle dynamics\cite{DynVLD,reply, nemir18}, moving  VLD from the channel core towards the  walls.  We move it  to RHS of  \Eq{mainEq}, such that after averaging of the core and walls regions, it will serve as an additional decay term in the channel core and as an additional production term near the walls. 
In such a way we get the ARD-type equation for the normalized VLD $L(x,t)$ for the core (labeled as '``c") and for the walls  (labeled as ``w") regions:

	\begin{eqnarray}\label{EqLx}\nonumber
	\frac{\partial L^j(x ,t)}{ \partial t} &+& \frac{\partial \C J^j(x ,t)}{\partial x} = 
	\tilde D^j \frac{\partial^2 L^j(x ,t)}{\partial x^2}+\tilde F^j[L( x,t)]\, ,\\
 \C J^j(x ,t)&=& V^0\sb s  L^j(x ,t)+\tilde {\C  J}^j_x(x ,t)\\\label{FL}
	\tilde F^j[L( x,t)]&=&F^j[L( x,t)]-\frac{\partial\C J^j_y(x ,t)}{\partial y},\quad j\in\{\rm c, \rm w\}\,.
	\end{eqnarray}
 The longitudinal tangle-induced flux $\tilde{\C J_x}=\C J_x-V^0\sb s L $ helps to redistribute the vortex line density along the tangle. 
  We account for it by replacing $V^0\sb s\to V^x\sb{s}$. Here we neglected  the streamwise component of the mutual friction contribution to the drift velocity $V^x\sb{mf}$  as  it contributes only about 1\% to the value of $V^x\sb{drift}$.  The modified ``reaction" term $\tilde F^j[L( x,t)]$  includes  the contribution from the transverse flux.
Since, in this formulation,  the effective diffusivity is a parameter that depends on the flow conditions, we allow for different values of $\tilde D^j$ for the channel core and for near-walls regions. Moreover, the values may differ in the tangle bulk and in the fronts regions.

Using this framework, we analyze in the rest of this Section various aspects of  the propagation of the  fronts, including the type of the fronts, their speeds, shapes, and the effective diffusivity. 

\begin{figure*}
		\includegraphics[scale=0.5]{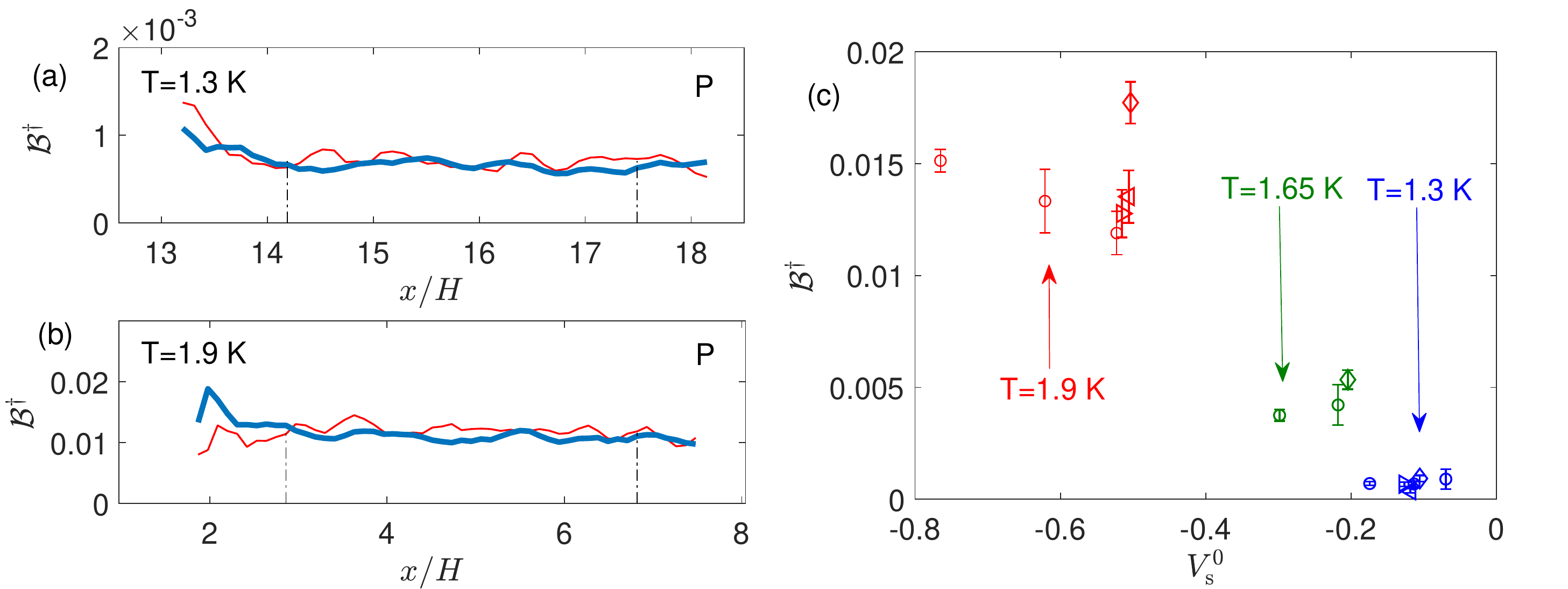}
	\caption{\label{f:Bdag}The coefficient $\C B^{\dagger} $ for various conditions. Streamwise profiles for (a)  $T=1.3$~K, $U_c=3$ cm/s   and  (b) $T=1.9$~K, $U_c=1$ cm/s. Thin red lines correspond to the core profiles, thick blue lines denote near-wall profiles.  Vertical dot-dashed lines  mark the edges of the tangle bulk. In calculation  of $\C B$ for the core and for the walls regions we used  the corresponding instantaneous values of $\beta, c^2_2$ and $ \C L_0$ and then averaged over time.  (c)  $\C B^{\dagger} $ averaged over tangle bulk. Symbols, denoting various flow conditions, are the same as in \Fig{f:c22all}.}
\end{figure*}

\begin{figure*}
		\includegraphics[scale=0.5]{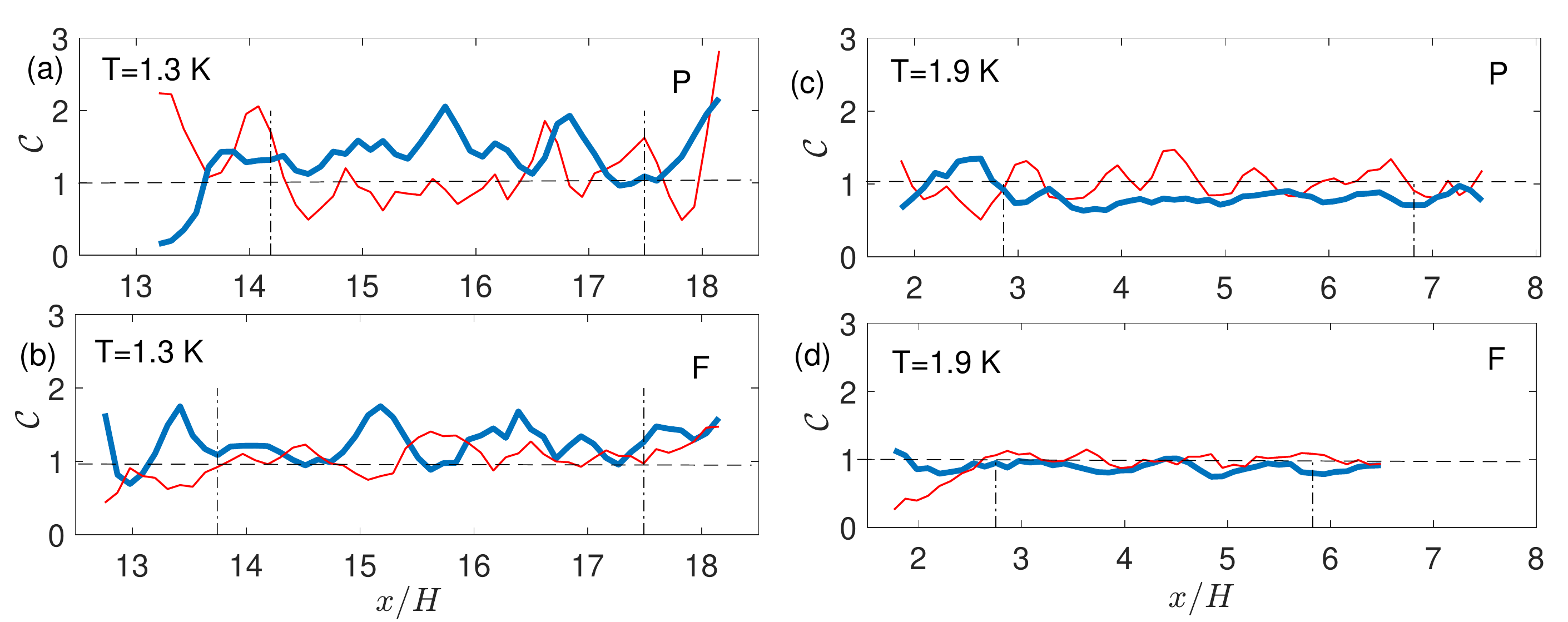}
	\caption{\label{f:16}The ratio $\C C $ for various conditions.  The profiles for the parabolic $V\sb n$  with (a) $T=1.3$~K, $U_C=3$ cm/s and  (c) $T=1.9$~K, $U_c=1$ cm/s.  The profiles of $\C C$ for the corresponding flattened profiles are shown in (b) for $T=1.3$\,K and in (d) for  $T=1.9$~K. Thin red lines correspond to the core profiles, thick blue lines denote near-wall profiles.  Vertical dot-dashed lines  mark the edges of the tangle bulk.  Horizontal dashed lines mark the value $\C C=1$. }
\end{figure*}

 \subsection{\label{ss:Ftheta}Properties of $\tilde F(L)$}
To identify the type of nonlinearity in the \Eq{EqLx}, we calculate the front profile for $\tilde F(L)$, as described in Appendix \ref{a:2}. The dependencies  of $\tilde F(L)$  and $\tilde F(L)/L$ on $L$ in the front regions , calculated for $T=1.3$~K, are shown in  \Fig{f:13}.   The results for the parabolic profile as shown by solid lines, for the flattened profile-- by dashed lines. As is clearly seen, the $L$ dependence of $\tilde F(L)$  is different for the hot fronts [panel (a)] and for the cold fronts [panel (b)]. The hot fronts are of the FKPP type, i.e the largest rate of growth $\sup_L [\tilde F(L)/L]$ is at $L \to 0$, as is shown in \Fig{f:13}(d), while for the cold fronts [\Fig{f:13}(c)], it is found closer to the center part of the front. This property is robust and observed all  flow conditions, and for both types of the $V\sb n$ profiles, although at $T=1.9$~K the maximum growth rate for the cold front is found closer to $L=0$ than at low $T$.
Despite complicated shapes of $\tilde F(L)$ for various flow conditions, the values of the largest rate of growth $\sup_L [\tilde F(L)/L]$, shown in \Fig{f:14}, depend linearly on $V^0\sb s$.  The dependencies  for the walls and the core regions differ even for the same front region, with hot fronts being stronger dependent on the advecting velocity than the cold fronts. In particular,   $\sup_L [\tilde F(L)/L]$ for the cold front in the channel core is almost $V^0\sb s$ independent.

Most of attempts to find equation of motion for the vortex line density so far dealt with  steady-state tangles and represented $\C P_1$ and $\C D$  in \Eq{terms} as functions of $\C L $ and $V\sb{ns}$ only for the homogeneous tangles, adding the curvature, the binormal, and their  derivatives\cite{Lipniacki01,JoiMonjovi06,aniso} in the inhomogeneous case.  In  the current situation of  the inhomogeneous and growing tangle we can not expect a unique closure. Aiming at the analysis of front dynamics, we make use of the fact  that at least the hot fronts are of FKPP type.  We then seek to represent \Eq{FL} in a general form
\begin{equation}\label{AB}
\tilde F(L)=\C A \, L-\C B \, L^2,
 \end{equation}
 where coefficients $\C A$ and $\C B$ have dimensions [1/s] and may depend on the position and time. 
 \subsection{\label{ss:model}Closure for $\tilde F(L)$}
 The main idea behind all the proposed closures\cite{schwarz88,DynVLD} is to take slowly-varying fields out of the average along the vortex lines. The resulting closure form is a product of slowly-varying macroscopic properties of the flow [such as $\B V\sb{ns}(x,y,t)$] and of the tangle [such as $c^2_2(x,y,t)$ and $I_{\ell,x}(x,y,t)$].  In Appendix \ref{a:3} we discuss various contributions to the $\tilde F(L)$.
  For our current analysis, however, we do not need all of them.
  
 We start with the last term in \Eq{AB}. It is readily associated with the term $\C D$, \Eq{decay},
as the relation $\C D\propto \C  L^2$  was shown experimentally\cite{Coex,DecayJETP} and rationalized theoretically\cite{Vinen,schwarz88} for the steady-state homogeneous vortex tangles. We use here the following form\cite{DynVLD} of this dependence for the dimensionless VLD $L$:
\begin{equation}\label{decaymod}
\C D\approx \alpha \beta \langle \varkappa^2\rangle L=\C B\,  L^2\, ;\quad \C B=\alpha \beta c_2^2 \C L_0 \, ,
\end{equation}
where the relation $ \langle \varkappa^2\rangle =c_2^2 \C L$  was used.
In the homogeneous tangle, where $c^2_2$ is a constant, the coefficient $\C B$ is also a constant for a given temperature. 
 In the inhomogeneous developing tangle, the mean-square curvature $ \langle \varkappa^2\rangle$ in the bulk of the tangle is almost homogeneous across the channel, while the vortex line density is not. Therefore, the coefficient $c^2_2$ has  more complicated behavior, as is  shown in \Fig{f:c22}. Nevertheless, when $c^2_2$ is averaged over the core and near-walls regions separately, the closure \Eq{decaymod} works quite well, especially at low $T$, as is shown in Appendix \ref{a:3}, \Fig{f:decay}.

 It turned out that the values of $\C B$ are very weakly dependent on the position in the channel. The difference between the values of $c^2_2$ in the channel core and near the walls is compensated by the corresponding difference in the values of $\C L_0$, such that $\C B$ is almost constant everywhere  in the channel,  with the exception of the immediate vicinity of the tangle edge, where the  measurements of $c^2_2$ become unreliable.  To compare  $\C B$ for various flow conditions,  we plot in \Fig{f:Bdag} the dimensionless 
 
\begin{equation}\label{Bdag}
\C B^{\dagger}=\C B \kappa/ (V^0\sb{ns})^2\approx \alpha\, c^2_2\, \Gamma^2\, ,
\end{equation}
 where in the right-most relation we took into account\cite{recon14} that $\ln(R/a_0)/(4\pi)\approx 1$ and $\Gamma=\kappa^2 \C L_0/(V^0\sb{ns})^2$ is a dimensionless coefficient relating\cite{recon14} the steady-state homogeneous VLD and the counterflow velocity. The coefficient $\C B^{\dagger}$ is expected to be a rising function of the temperature, but to have  only weak dependence on other flow conditions. The  streamwise profiles of $\C B^{\dagger} $ are illustrated  for $T=1.3$~K, $U_c=3$ cm/s (\Fig{f:Bdag}a) and   $T=1.9$~K, $U_c=1$ cm/s (\Fig{f:Bdag}b). The values of    $\C B^{\dagger}$ averaged over tangle bulk are summarized in \Fig{f:Bdag}c.  As expected, the coefficients $\C B^{\dagger}$  grow with the temperature, but otherwise, despite some scatter, are almost independent of the flow conditions.  Note that, in accordance with the behavior of $c^2_2$, $\C B^{\dagger}$ is larger in the flow generated by the flattened $V\sb n$ profile (diamonds), that in the flow, generated by the corresponding parabolic profile (circles), for similar $V^0\sb s$ .

Using almost constancy of $\C B$  over entire tangle, we can associate $ \tau\sb {dec} \equiv (\C B)^{-1}$ with some characteristic time, in this case of the tangle decay, and  further rewrite 
\begin{equation}
\tilde F(L)=\frac{ L}{\tau\sb {dec} }(\C C-L)\, ,\quad \C C=\frac{\C A}{ \C B }\, .
\end{equation}
In the steady-state homogeneous tangle, $\C C =1$. Again, there is no \emph{a-priori} reason to expect that this relation will hold in the current situation. However, as is shown in \Fig{f:16},  up to natural fluctuations, $\C C\approx 1$  along all the tangle including front regions, with accuracy about 20\% -30\% depending on the stage of the tangle development.  In particular, at $T=1.3$\,K the near-wall regions are more dissipative that the channel core, while at high $T$ they are less dissipative.
The closeness of the ratio $\C{C}$  to unity indicate that  we correctly account for all the relevant contributions to the $\tilde F(L)$ in  \Eq{FL}.
\begin{figure*}
	\includegraphics[scale=0.42]{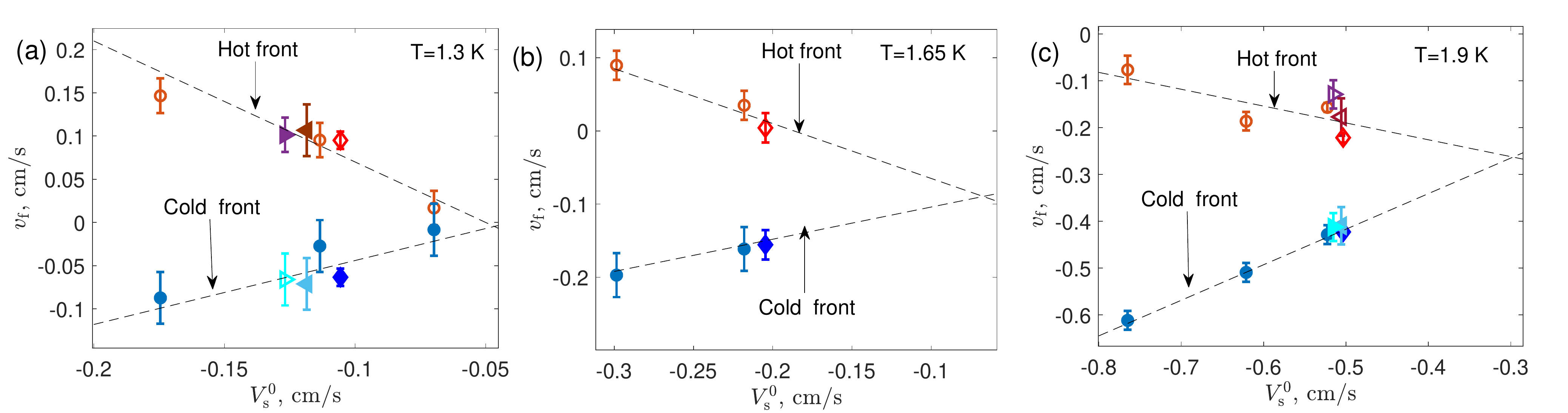}	
	\caption{\label{f:12} Front velocities as a function of mean superfluid  velocity for (a) $T=1.3$~K, (b) $T=1.65$~K and (c) $T=1.9$~K. In all panels, different flow conditions are represented by different symbols: $\circ$ denote front velocities for the  parabolic $V\sb n$ and various $U_c$, $\diamond$ corresponds to the flattened $V\sb n$ profile, $\triangleright$  and $\triangleleft$ denote channel width $H=0.15$ cm and  $H=0.2$ cm, respectively. The symbols, marking cold front velocities, are filled, the symbols for the hot front velocities are empty. Dashed lines serve to guide the eye only. }
\end{figure*}

\begin{figure*}
		\includegraphics[scale=0.44]{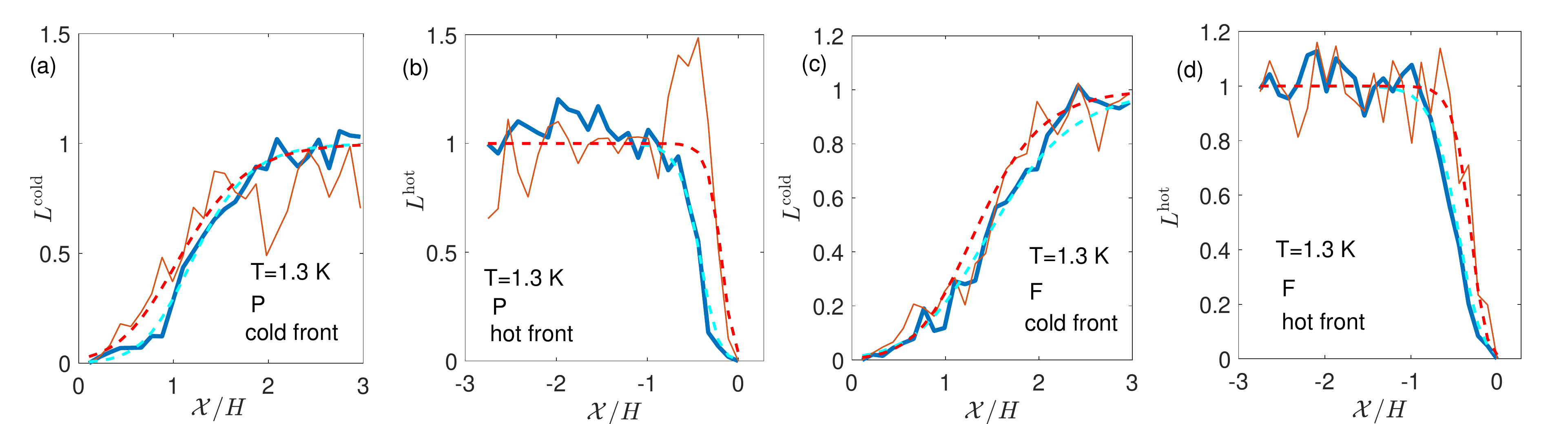}\\
		\includegraphics[scale=0.44]{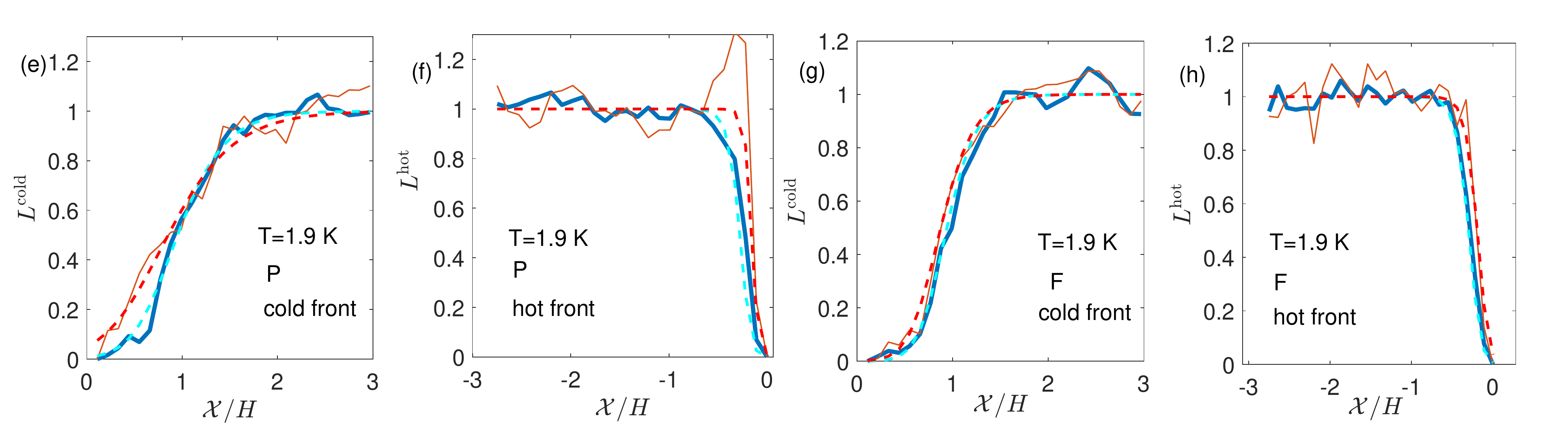}
	\caption{\label{f:17}The cold and hot fronts shapes for various conditions: (a,b) $V\sb n$  with $T=1.3$~K, $U_C=3$ cm/s ; (c,d) $T=1.3$K and flattened $V\sb n$ profile; (e,f)  $T=1.9$~K, $U_c=1$ cm/s; (g,h) $T=1.9$K and flattened $V\sb n$ profile. The cold front shapes are shown in panels (a,c,e) and (g), the hot front shapes ares shown in panels (b,d,f) and (h). Thin red lines correspond to the core fronts, thick blue lines denote near-wall fronts.   Cyan dashed lines denote fits for the near-wall fronts, red dashed lines denote fits for the  fronts in the channel core. }
\end{figure*}
\subsection{\label{ss:sol} Solution of VLD equation of motion}
Having defined the functional form for $\tilde F(L)$ and taking, for now, $\C{C}=1$, we can return to \Eq{EqLx} and  rewrite it as

	\begin{eqnarray}\label{EqLxL}
\partial_t L^j(x ,t)&+&V^x\sb{s} \partial_x L^j(x,t)= \\ \nonumber	\tilde D^j  \partial_{x,x} L^j(x ,t)&+&1/\tau \sb{dec}~ L^j(x,t)  [1-L^j(x,t)]\, .
\end{eqnarray}
We now switch to dimensionless variables (omitting for shortness the  index $j$)
\begin{eqnarray}\label{dimless}
\tau&=&t/\tau\sb{dec},\quad  z=x/\sigma,\quad \sigma=\sqrt{\tilde D \tau\sb{dec}},\\
 w&=&V^x\sb{s} /V\sb{diff},  \quad V\sb{diff}=\sigma/\tau\sb{dec}\, ,
\end{eqnarray}

 to rewrite \Eq{EqLxL} as
\begin{equation}\label{ARDL}
\partial_{\tau} L^j+w^j ~\partial_z L^j=\partial_{z,z}L^j+L^j-(L^j)^2\, .
\end{equation}
Comparing with \Eq{ARDE},  we see that \Eq{ARDL} is the ARD equation of FKPP type for the vortex line density, which for front velocities $v\sb f>2\sqrt{\tilde D/\tau_d}$ admits a traveling wave solution $\zeta=c(z- V\sb f \tau)$ with the dimensionless front speed $V\sb f=v\sb f/V\sb{diff}$. Substituting this solution to \eqref{ARDL}, we get an equation that defines the  velocity and  the shape of the front:
\begin{eqnarray}\label{finalEq}
[c_j \,  v^j\,   \partial_{\zeta}  &+&c^2_j \, \partial_{\zeta, \zeta}]L^j +L^j -(L^j)^2=0\, ,\\ \nonumber
\quad  v^j&=&V^j\sb f-w^j\, .
\end{eqnarray}

A similar equation was obtained by Nemirovskii\cite{nemir11} for 1D front propagation, using the original Vinen's form for $F(L)=\alpha_{Vi}L^{3/2}-\beta_{Vi}L^2$,  and solved numerically for the front speed, with 
the parameters estimated for the homogeneous steady-state vortex tangle by Schwarz\cite{schwarz88} and the diffusion constant\cite{nemir10} $D\approx 2.2 \kappa$.

The equation \eqref{finalEq} may be solved analytically using Tahn method\cite{Tanh1,Tanh2}. A general form of these solutions,
symmetric with respect to the direction of propagation, reads:
\begin{equation}\label{FS}
L(\zeta)=\frac 14 \big [1\pm \tanh \zeta \big]^2\,, \ c=\frac1{2\sqrt 6}\,,\ \  v= \mp \frac5 {\sqrt 6}\, ,
\end{equation}
or, relaxing the requirement that $\C{ C}=1$, 
\begin{equation}\label{FSAB}
L(\zeta)=\frac {\C C}{4}\big [1\pm \tanh \zeta \big]^2\,, \ c=\frac{\sqrt{\C{C}}}{2\sqrt 6}\,,\ \  v= \mp \frac{5\sqrt{\C{C}}} {\sqrt 6}\, ,
\end{equation}
Returning to the original dimensional variables 
\begin{eqnarray}
L(x)&=&\frac14 \big [1\pm \tanh (\frac{1}{\lambda}[x-v\sb f\,  t] )\big]^2\, ,\\\label{Fwidth}
\lambda&=& 2\, \sigma\sqrt{6/ \C C}\, , v\sb f =\pm 5\,V\sb{diff} \sqrt{\C C/6} +V^x\sb{s}\, ,
\end{eqnarray}
where $\lambda$ is the front width. As we can see, the effective diffusion constant and the characteristic decay time define both the front width and the front velocity via the diffusion spread $\sigma$ and its speed $V\sb{diff}$.

The similar (symmetric) solution was postulated in \Ref{vanBeelen88} without derivation, assuming $F(L)$ based on Vinen's form  of $F(L)$ for the case of thermal counterflow\cite{Vinen3} in the presence of a wall.

However, as we know now, the hot and cold fronts are of different types.  Strictly speaking,  only hot fronts are of FKPP type (pulled) and fulfill the underlying assumptions for the solution. Nevertheless, we may hope that,  at least at high temperatures, the solution will describe reasonably well  also the cold fronts.

Recalling that $\tilde D$ depends on the flow conditions and therefore may be different for the channel core and near the walls, we get solutions for four fronts:
\begin{eqnarray}\label{FS2a}
L^{j,\rm c}(x)&=&\frac14 \big [1+ \tanh (\frac{1}{\lambda^{j,\rm c}}[x-v\sp{c}\sb f\,  t] )\big]^2\,,\\\label{FS2b}
L^{j,\rm h}(x)&=&\frac14 \big [1- \tanh (\frac{1}{\lambda^{j,\rm h}}[x-v\sp{h}\sb f\,  t] )\big]^2\,,
\end{eqnarray}  
where $\lambda^{j,\rm c},\lambda^{j,\rm h}$ are the  widths  of the corresponding fronts and $v\sp{c}_f,v\sp{h}_f$  are the corresponding front velocities.
Here a word of caution is in order. The solutions \Eq{FS2a}-\eqref{FS2b} do not describe any transient behavior, such as VLD hump in the channel core near the hot front, strong VLD fluctuations at the fronts at low $T$, or effects of the different type of the nonlinearity for the cold fronts. Since fronts of the studied tangles most probably did not reach the expected limiting shapes, all parameters are considered as effective and corresponding to the chosen time $t\sb f$.

 The mean front shapes were calculated using the procedure described in Appendix \ref{a:2} and fitted with the solutions \Eq{FS2a}-\eqref{FS2b} to obtain the front velocities $v\sp{h}\sb f,v\sp{c}\sb f$ and front widths $\lambda^{j,\rm{c}},\lambda^{j,\rm{h}}$. 

 \subsection{\label{ss:vf}Front  velocities and shapes}

The front speeds are shown in \Fig{f:12} as a function of the mean superfluid velocity $V^0\sb s$.  It is clearly seen that $v\sb f$ depends linearly on the advecting velocity, with hot and cold front speeds having opposite trends, independent of the actual orientation of the hot front velocity. All data for a given temperature are well fit by the same linear dependence, shown as  black dashed lines.  Note that the front velocities are the same for the channel core and the near-walls regions. This point requires additional attention.  As we mentioned earlier,  the hot front is lead by the channel core, while the cold front is defined by the near-wall region. Moreover, the superfluid velocity in the bulk of the channel, as we showed in \Fig{f:8}, is close to the corresponding $v\sb f$. This raises a natural question, how the hot front velocity near the channel walls become equal to that in the core and similarly, the cold front velocity  in the core becomes equal to $v\sp c\sb f$ near the channel walls? The answer lies in the action of the transverse flux $\partial \C J^j_y(x ,t)/\partial y$ that changes very strongly in the fronts regions, but is almost constant along the tangle, see \Fig{f:prodX} and  Appendix \ref{a:3}.
In this way, the hot front near the wall  is formed by VLD brought by the flux from the channel core and its velocity matches the velocity at the core only very close to the tangle edge. Similarly, the superfluid velocity in the core of the channel is quickly changed to $v\sp c\sb f$ by the transverse flux which in this region brings VLD from the walls toward the channel core.  Here, the flux is much weaker than in the hot front region and the development of the cold front in the channel core is a result of a complicated interplay of various mechanisms,  leading to long-lasting  transient behavior.

 At low $T$, $|v\sp h\sb f|>  |v\sp c\sb f|$ for the same advecting velocity $V^0\sb s$, while at high $T$ the relation is opposite. This observation is in agreement with the early experiments is thin capillaries\cite{PeshkovTkachenko, Tough}. Moreover, the front velocities, observed in \Ref{PeshkovTkachenko} for $T=1.34$ K at low  heat fluxes, are similar to  $v\sb f$ measured in our simulations at $T=1.3$ K. The linear dependencies point out to a particular value of $V^{0*}\sb s$ at which the front speeds are expected to be the same and, therefore, only one front can propagate.  It is natural to associate the corresponding $v^{*}\sb{f}$ with the onset of the front solution in the counterflow. Since without  advecting flow (or more specifically, the counterflow $V\sb{ns}$),  the counterflow turbulence does not exist, the  onset front velocity $v^*\sb f>v_0$.   Note that at all temperatures, the values of  $V^{0 *}\sb{s}$ are larger than the critical $V^0\sb{s,c}$, below which the vortex tangle is not formed. The values of  $V^0\sb{s,c}$ were estimated from the $\sqrt{\C L}=\gamma(V\sb{ns}-v_c)$ dependence and the counterflow condition.
 The onset fronts velocities $v\sb f^*$ and  $V^{0*}\sb{s}$ are listed in the Table \ref{t:3}.

\begin{table}[t]
	\caption{\label{t:3} Onset front velocities 	$v_f^{*}$  and corresponding mean superfluid velocities $V^{0,*}\sb{s}$. The error-bars reflect the sensitivity of the linear fitting procedure. 
	}
	\begin{tabular}{   c|c c c }
		\hline
		$T$,  K    ~~~         	&~~~~1.3~~~~ & ~~~~1.65~~~~&~~~1.9~~~~~~  \\ \hline
	$v^*\sb f$,\, cm/s 	~~~&  $ -0.005\pm 0.02$     &   $-0.09\pm0.02$      & $-0.26\pm0.02$  \\
	 $V^{0*}\sb{s}$,\, cm/s 	~~~ &   $-0.05\pm0.02$     &  $-0.07\pm 0.02$         &$-0.30\pm0.02$\\ \hline
	\end{tabular}
\end{table}

The representative front shapes, together with their fits with the solution \Eq{FS2a}-\eqref{FS2b},  are shown in \Fig{f:17}.
The $x$-axis shows the distance from the front edge ($\C X=0$)  for the core and the walls regions separately. First of all, we note the presence of the  narrow VLD hump, localized between the tangle bulk and the hot front in the tangles driven by the parabolic $V\sb n$ profiles, \Fig{f:17}a,c. This hump is not  formed when the normal-fluid velocity profile is flattened, \Fig{f:17}b,d. In all cases, the hot fronts are  $2-5$ times more narrow than the cold fronts. The hot fronts are steeper  in the core of the channel, than near the walls, while  cold fronts are steeper  near the walls, or similar. The presence of a shallow shoulder at small $L$ in the cold front shapes, well seen for the near-wall front shapes  at both temperatures, is a sign of non-FKPP non-linearity and  is not accounted for by the solution.  However, the solution \Eq{FS2a} describes reasonably well the overall cold front shapes, especially at high temperatures, at which the fronts are well-formed and developed.

\subsection{\label{ss:Deff} Effective diffusivity}

The importance of the diffusion mechanism for the decay of inhomogeneous tangle was studied  theoretically\cite{nemir10} and  numerically \cite{tsubota03,GPEDiff,EffDiff3D} for the decaying tangles at $T=0$~K with most recent estimates of the effective diffusion constant in the range $(0.1-1)\kappa$. The presence of dissipative walls reduces\cite{GPEDiff} the values of the effective diffusion constant, while in the 3D unbounded vortex tangle\cite{EffDiff3D} the value of the effective diffusion constant was found to be close to $0.5\kappa$.

Using the relation between the front width and the effective diffusion constant, \Eqs{dimless} and \eqref{Fwidth}, we can estimate $\tilde D$ for various conditions. 
 For that, we rewrite \Eq{Fwidth} as
 \begin{equation}\label{Dtilde}
 \tilde D^{j,\rm{c}}=\frac{ (\lambda^{j,\rm{c}})^2}{24 \tau_d}\, ,\quad  \tilde D^{j,\rm{h}}=\frac{ (\lambda^{j,\rm{h}})^2}{24 \tau_d}\, ,
 \end{equation}
 where we  retain $\C C=1$ and $\tau_d=\rm{Const.}$ for given conditions. To get an idea of what behavior to expect from  $\tilde D$ we rewrite \eq{Dtilde} (omitting indices $j$, c and h for clarity) as
 \begin{equation}\label{Dest}
 \tilde D=\frac{ \lambda^2 \C B}{24}=\frac{ \lambda^2 \C B^{\dagger} (V^0\sb{ns})^2}{24 \kappa}\,.
  \end{equation}
  The temperature dependence of $\tilde D$ is therefore mostly defined by $\C B^{\dagger}$, the dependence on the driving velocity by $(V^0\sb{ns})^2$ and the influence of other flow conditions, including the spatial dependence -- by the front width $\lambda$. There is no systematic dependence of the front width on the driving velocity. Recall that the cold fronts are wider than the hot fronts, such that for the given $T$ and $V^0\sb{ns}$, $ \tilde D^{j,\rm{c}}>\tilde D^{j,\rm{h}}$ with the difference reaching up to an order of magnitude. The typical  front width range decreases with temperature, such that  $\lambda \sp c\sim(0.05-0.1)$\,cm at $T=1.3$\,K, while at $T=1.9$\,K, $\lambda \sp c\sim(0.03-0.05)$\,cm. The  hot fronts are more narrow:  $\lambda \sp h\sim(0.02-0.05)$\,cm at $T=1.3$\,K, while $ \lambda \sp h\sim(0.005-0.03)$\, cm at  $T=1.9$\, K. On the other hand, $\C B^{\dagger}$ grows with $T$.  As a result, for the studied range of flow conditions, at the cold front, the typical values  are $\tilde D\sp c\sim(0.5-1.5)\kappa$, while at the hot fronts $\tilde D\sp h\sim(0.01-0.1)\kappa$ and are larger for higher temperatures. This $T$-dependence is more prominent for the flows driven by the parabolic normal-fluid velocity. The representative values of $\tilde D$, calculated according to \Eq{Dtilde}, are listed in Table \ref{t:4}. It is important to remember that the effective diffusivity $\tilde D$ is not a material property of superfluid $^4$He, but a dynamical property of propagating fronts in the particular flow conditions, including different nonlinear processes in the front regions. In addition, the values listed in the table correspond to the  reached stage of the tangle development and are  sensitive to the presence of the transient processes in the tangle core.  Nevertheless, since the order of magnitude of $\tilde D$ is the same for the flows driven by the parabolic  and by flattened $V\sb n$ profiles at all studied temperatures, these values may be considered as a robust dynamical property of the propagating fronts in the channel counterflow. 
  
  The values of $\tilde D$  at the hot front are remarkably close to the values of the effective diffusion constant found numerically in the bounded\cite{tsubota03,GPEDiff} and unbounded\cite{EffDiff3D} bulk tangles at zero temperature.
  We do  not have a reliable measure of the diffusion in the  bulk of the tangle. However, since the values of many of the tangle properties  in the bulk  are similar to those in the hot front region, we  suggest  that also the values of $\tilde D$ in the tangle bulk would be similar to those in the hot front region at least in the order of the magnitude.

 \begin{table*}
 	\begin{tabular*}{\linewidth}{@{\extracolsep{\fill} } cc c c c c c}
 		\hline 
 		&    \multicolumn{2}{c}{$T=1.3$ K} &   \multicolumn{2}{c}{$T=1.65$ K}  & \multicolumn{2}{c}{$T=1.9$ K} \\ 
 	Type $V\sb n(y)$	& P & F & P & F & P &F \\ \\
 	$\tilde D\sp{core,c}/\kappa$	&$ 0.5\pm 0.2$ & $1.4\pm 0.5$& $0.7\pm 0.3$ &$ 1.1\pm 0.3$ & $1.3\pm 0.4$&$1.3\pm0.3$ \\ 
 		$\tilde D\sp{wall,c}/\kappa$	& $0.4\pm 0.1$&$1.0\pm0.5$&  $0.8\pm 0.3$& $0.7\pm 0.2$ &$1.3\pm 0.3$ &$1.3 \pm0.3$\\ 
 		$\tilde D\sp{core,h}/\kappa$	& $0.01\pm0.005$ &$0.02\pm0.01$ &$0.06\pm 0.02$&$ 0.05\pm 0.02$ &$ 0.04\pm0.01$&$0.08\pm0.02 $\\ 
 		$\tilde D\sp{wall,h}/\kappa$	&$0.03\pm 0.01$& $0.12\pm0.04$& $0.05\pm0.02$& $0.11\pm 0.01$ &$0.20\pm0.05$
 	& $0.32\pm0.07$ \\ 
 		\hline 
 	\end{tabular*} 
 		\caption{\label{t:4}Effective diffusivity at the cold (c) and hot (h) fronts near the wall and in the channel core  for representative conditions. "P" and "F" denote the parabolic and the flattened  $V\sb n$ profile, respectively. The error-bars account for $\C C=1\pm0.2$ as well as the errors in measurements of $\lambda$ and $\C B$. The flow conditions are the same as in \Fig{f:c22} and \Fig{f:ilx}.}
\end{table*}

 \section{\label{s:Discussion} Discussion}
Our simulations of the quantum vortex tangles that develop freely in the channel from localized initial conditions under the influence of the counterflow velocity, give a unique insight  into their natural dynamics and structure. Despite a wide variety of the flow conditions experienced by the vortex lines  that influence the local dynamics, there are many common features.

 In particular, the tangles may be divided into regions according to their dynamics. The regions near the tangle edges exhibit front dynamics.  The dynamics of tangle bulk is  more similar to that of the steady-state stationary tangles.  In the bulk, the parts of the tangle that develop near the channel wall, are first to reach  equilibrium  VLD and grow almost symmetrically with respect of the direction of the counterflow velocity. On the other hand, the transient tangle dynamics  in the channel core is slower, with notable asymmetry and preferential growth of VLD toward the hot front, resulting in the long-lasting streamwise inhomogeneity. This behavior is similar at high and low temperatures, despite the different direction of the hot front propagation. This asymmetry is originated from the production of the vortex  line length, strongly peaked  in the channel core within the hot front region. The only difference between the  dynamics at different  velocities of the driving normal fluid and even its wall-normal profile is the duration of the transient behavior and degree of the inhomogeneity of resulting vortex tangle. Conversely, the structural properties of the vortex tangle, such as the ratio between the curvature and the vortex line  density and preferential orientation of the local velocity,  reach their steady-state distributions as soon as the tangle become three-dimensional, with core values similar to those obtained in the simulations of the steady-state vortex tangles and  the experimental estimates. 

The VLD is higher near the walls than in the channel core, peaking at about the intervortex distance, in agreement  with the results of simulations of steady-state tangles in the channel. This difference between the channel core and the near-wall regions is less prominent when the flow is driven by the normal-fluid velocity with the flattened profile. A similar trend of relatively flat VLD distribution in the channel core, that extends towards the walls, was observed in simulations with wider channels at all temperatures.

An explicit account for the advecting mean superfluid velocity allowed us to detect a superfluid motion of various scales within the vortex tangle. The largest scales of this motion reach the channel size at strong driving velocity.  When normal-fluid velocity profile is flattened, as is expected in the turbulent flow, superfluid motions exist at many scales. The presence of this large-scale superfluid motion is reflected in the streamwise inhomogeneity of various  tangle properties. The typical period of the fluctuations  is of the order $H/2$, corresponding to the largest eddies  formed in the tangle.

The analysis of the dynamics of the fronts  in the framework of the advection-diffusion-reaction equation gives unexpected results. The two fronts are driven by different parts of the flow and have a different type of nonlinearity of the generalized production term. The hot fronts are ``pulled", i.e. driven by the flow in the  channel core and the leading edge dominate in defining their  high steepness and the propagation speed.   The cold fronts, on the other hand, are lead by the near-walls tangle and are ``pushed" by the nonlinear interior. A low  density  "foot" moves before the tangle, and only at about quarter of the front width the VLD start to rise fast. These fronts are wide and the shape difference between the channel core and near the walls is larger.  In accordance with ADR dynamics, the front velocities are linearly proportional to the advecting mean superfluid velocity, with common dependence for all conditions at a given temperature. The  analytic solution of the equation of motion \Eq{EqLx} fits well the overall front shapes for all conditions, while it does not describe the transient effects near the hot fronts and the effects of the non-FKPP nonlinearity at the cold fronts. These solutions allow extracting the effective diffusivity which is flow-dependent and different at the hot and at the cold fronts. The values  of the effective diffusivity measured the hot fronts agree in the order of magnitude with recent estimates from simulations at $T=0$\, K.

\appendix 

 \section{\label{a:1}  Calculation of various profiles.}
\begin{figure*}
 	\includegraphics[scale=0.6]{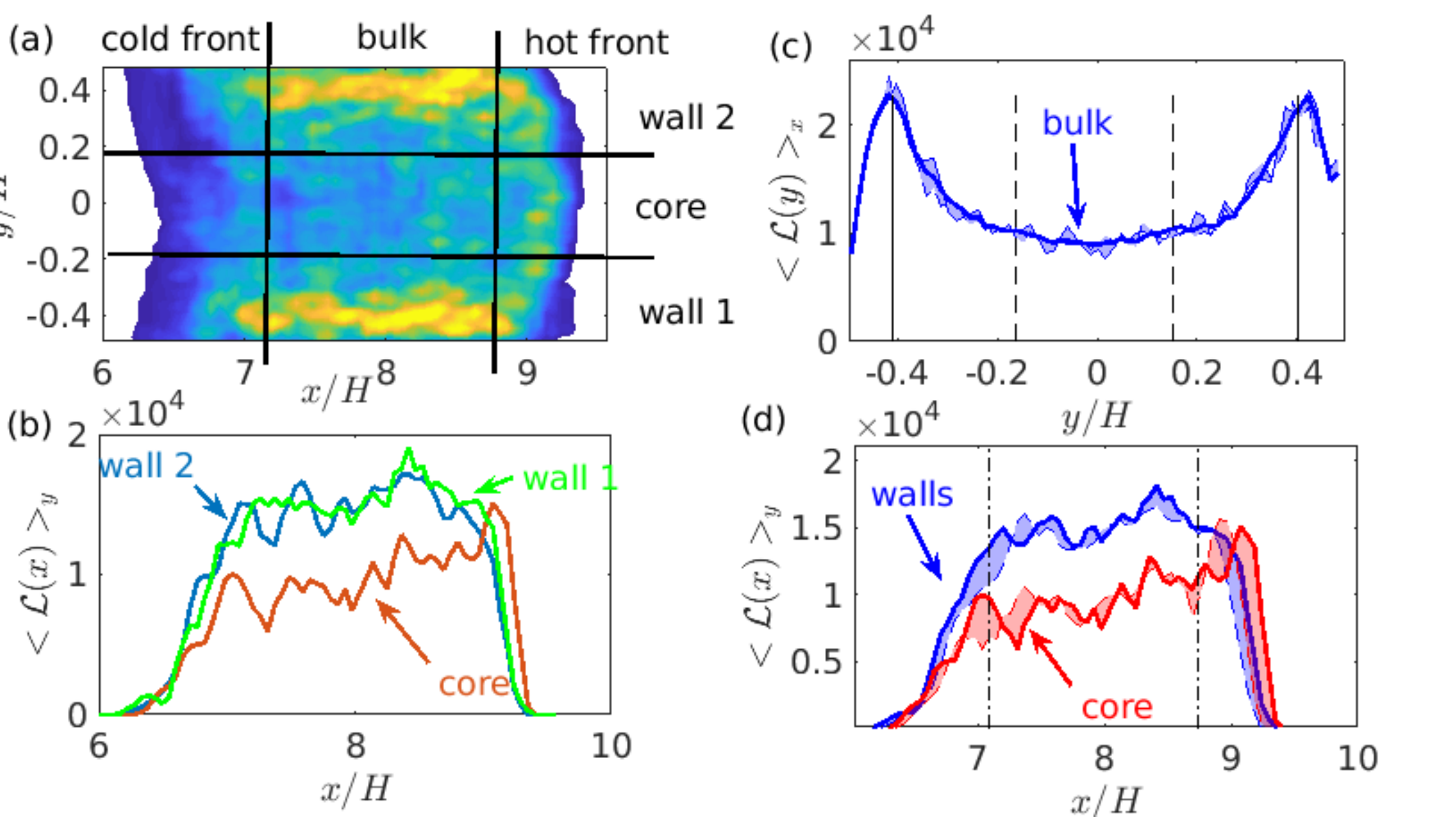}
 	\caption{\label{f:diagram} Schematic representation of various averaging zones. (a)  2D map of $\C L(x,y)$ (cm$^{-2}$) in which  various averaging zones are marked. (b) streamwise profiles $\langle \C L(x) \rangle_y$ averaged over two near-wall zones and over the core in the $y$ direction.  (c) The wall-normal profile $\langle \C L(y)\rangle_x$ averaged over the tangle bulk in the $x$-direction.  (d) the streamwise profiles  $\langle \C L(x) \rangle_y$, in which two near-wall zones are averaged together. In panels (c) and (d) the shaded area shows variation of VLD between $t\sb f-t\sb {av}$ (thin dashed lines) and $t\sb f$ (think solid lines).}
 \end{figure*}

 The wall-normal and the streamwise profiles of various quantities are calculated according to the scheme shown in \Fig{f:diagram}. The division into different zones is somewhat arbitrary, however, we have checked that the values of the tangle properties  are robust  with respect to the variation of the  zones boundaries within 2 mesh-sizes.  For illustration we use the vortex line density $\C L$. The wall-normal profiles were obtained by averaging the 2D maps over the bulk region of the tangle  defined at each time moment and further averaged over last $t\sb{av}=0.2$ sec. The shading in \Fig{f:diagram}(c,d) illustrates the variation  between the profiles at time $t\sb{f}-t\sb{av}$ ( dashed lines) and $t\sb{f}$ (solid lines).   The streamwise profiles were calculated for $t\sb{f}$ by averaging  over the core and  near-walls regions separately. In cases where the behavior at two near-walls regions was similar, they were averaged together. 
  The streamwise  profiles of structural properties, such as $c^2_2$, and various terms of the balance equation, in addition to averaging over core and near-walls regions, were averaged over last $0.1$ s of time evolution. In cases that involve division by $\C L$,  the points near the edge of the tangle, where $\C L(x)\approx 0$, were omitted in calculation of the time average and not shown.
 
 The intervortex  distance $\ell=\C L^{-1/2}$,  shown  in the wall-normal $y$-profiles as  vertical thin black lines, is calculated here at time $t\sb{f}$ by averaging $\C L$ over bulk in $x$-direction and over near-walls region in $y$-direction.

  \section{\label{a:2}  Front shape.}
\begin{figure*}[htp]
  		\includegraphics[scale=0.43]{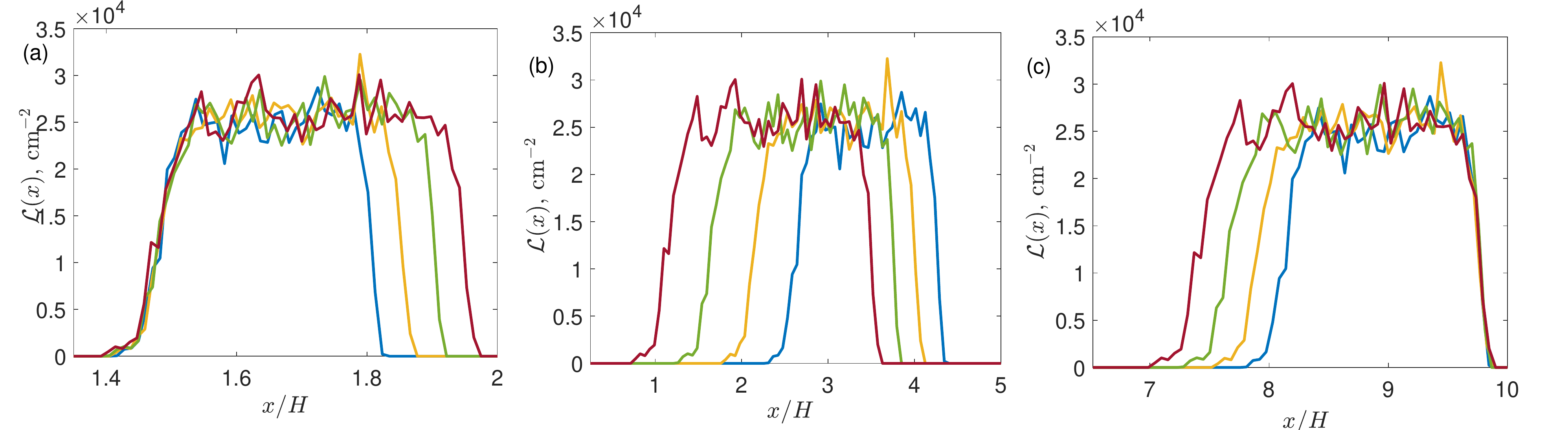}
	\caption{\label{f:collapse} The hot and cold front shapes. A series of streamwise VLD profiles corresponds to last 1 s of the evolution of the walls region, $T=1.9$~K, flattened $V\sb n$. (a) The profiles are collapsed using the cold front speed, (b) the original profiles, (c) the profiles are collapsed using the hot front speed.} 
 \end{figure*}

  The fronts of the tangles propagate without shape change. To show this, we shift the $x$-positions of  the streamwise VLD profiles $\C L(x)$, corresponding to the time period when the bulk and the fronts are fully developed, to the left and to the right, such that the corresponding  tangle edges overlap. This procedure is used to measure the front speeds $v^{\rm c}_f$ and $v^{\rm h}_f$ that allow such an overlap. The original profiles $\C L(x)$ are shown in \Fig{f:collapse}b. The result of the cold front collapse  is plotted in panel(a) and of the  hot front collapse-- in panel(c). Clearly, the front shape does not change during this time period. To obtain the front shape we calculate the dimensionless VLD $L=\C L/\C L_0$, where $\C L_0$ is the mean VLD in the bulk of the tangle. Since the values of VLD differ in the core of the channel and near the walls, we treat these regions separately.
    We further average these profiles over the time period of 0.2 s.  In such a way we obtain four shapes, for the cold front for the hot front in the core and in the near-walls regions, shown in \Fig{f:17}.

  The same procedure was used to obtain the front shapes of other quantities of interest.
 \section{\label{a:3}  Terms of balance equation.}
\begin{figure*}[htp]
  \includegraphics[scale=0.4]{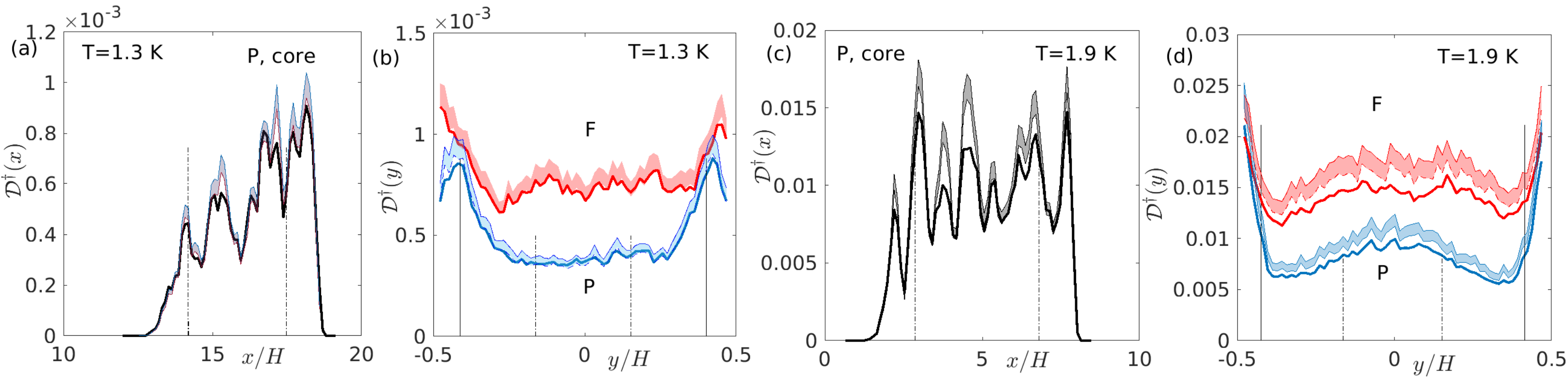}
  \caption{\label{f:decay} The decay term \Eq{decay} (thick lines) and its model form \Eq{decaymod} with 95\% confidence interval  (shaded area) at different conditions. (a,c) The streamwise profiles for the channel core  for (a) $T=1.3$~K, parabolic $V\sb n$ with $U_c=3$ cm/s and (c) $T=1.9$~K, parabolic $V\sb n$ with $U_c=1$ cm/s. (b,d)  The wall-normal profiles for the  conditions of (a) and (c), respectively, and matching flows with flattened $V\sb n$ profile. 
  	Dot-dashed black lines mark the edges of the bulk and the core regions for the streamwise and for the wall-normal profiles, respectively. Thin solid lines in panels (b),(d)  are placed at the intervortex distance from the corresponding walls. The profiles are calculated as described in Appendix \ref{a:1}. For normalization in \Eq{decay}, $\C L\sp{core}_0$ was used for  the streamwise profiles in (a) and (c) and  $\C L\sp{wall}_0$ for the wall-normal profiles in (b) and (d).}
\end{figure*}

 In this section, we provide a detailed description of various contributions to $\C F(L)$  used in the analysis of the front dynamics.
 As was shown in \Sec{ss:model}, the spatial distribution of the decay term $\C D\approx \alpha \beta c^2_2 \C L_0 L^2$ essentially follows $L^2$. This representation faithfully describes the integral form \eq{decay} not only on average in the steady-state tangle but also locally and instantaneously, including the transient stage of the dynamics,  as is shown in \Fig{f:decay}.  To allow comparison, the dimensionless values $\C D^{\dagger}=\C D \kappa/  (V^0\sb{ns})^2$ are plotted.
The model slightly overestimates the decay term at high $T$, but otherwise should be considered very adequate everywhere in the tangle. Note that the strong streamwise  inhomogeneity, amplified compared to VLD, is well reproduced by the model.
The situation is different with the production term. Directly interpreting the  model form as a product of average slowly-varying fields, we get for $\C P_1=\alpha \langle V^x\sb{ns,nl}\rangle \langle \ s^{\prime}\times s^{\prime \prime}\rangle _x L \approx \alpha V^0\sb{ns} I_{\ell,x}\, L$.  So far, the problem of the closure for $\C P_1$ amounted  to the question how to describe\cite{Vinen3,schwarz88,DynVLD,reply,nemir18} $ I_{\ell,x}$ in terms of $L$ and $V\sb{ns}$. As it follows from the discussion in \Sec{ss:dyn} and \ref{ss:c22}, in the inhomogeneous flows, there is no simple answer to this question. Additional complication arises at low $T$, at which  the contributions of $\C P_2=\langle \B s^{\prime}\cdot \B V^{\prime}\sb{nl  }\rangle\, L$ and $\C P_3=-\alpha^{\prime} V\sb{ns}\langle \varkappa\rangle  \,L$ near the walls are  not negligible.   We do not attempt here to find the best model representation, but rather  point out additional difficulties brought up by the presence of  large-scale superfluid motion.

\begin{figure*}[htp]
 	\includegraphics[scale=0.45]{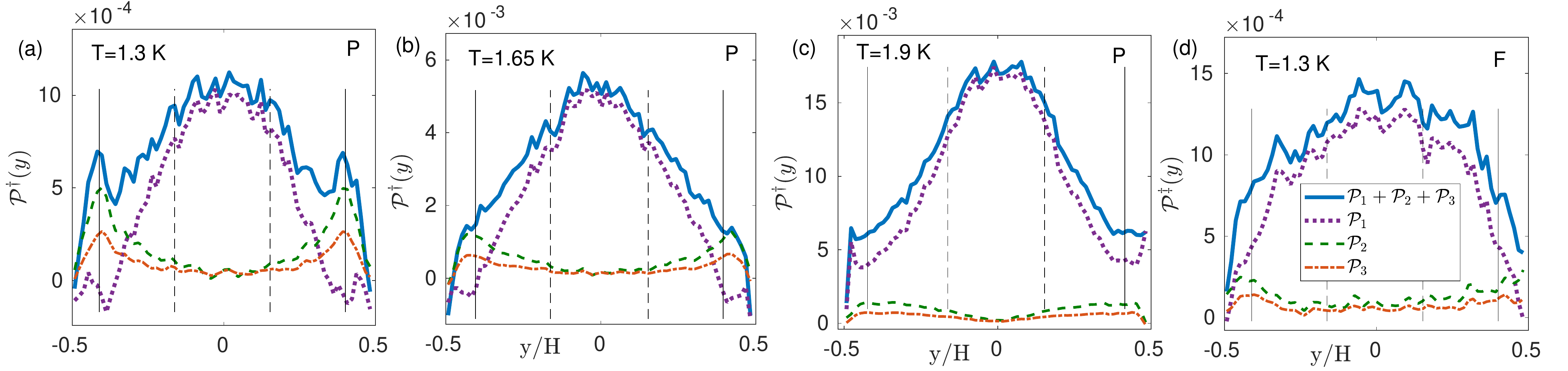}
 	\caption{\label{f:prod3} The  production terms:  $\C P_1$ (\Eq{p1}, purple dotted  line),   $\C P_2$ (\Eq{p2}, green dashed line),  $\C P_2$ (\Eq{p3}, brown dot-dashed  line)  and their sum (blue solid line)   at different conditions.   (a) $T=1.3$~K, parabolic $V\sb n$ with $U_c=3$ cm/s, (b) $T=1.65$~K, parabolic $V\sb n$ with $U_c=1.5$ cm/s , (c) $T=1.9$~K, parabolic $V\sb n$ with $U_c=1$ cm/s, (d) $T=1.3$~K, flattened $V\sb n$ profile.	Dashed black lines mark the edges of the core region. Thin solid lines are placed at the intervortex distance from the corresponding walls. }
 \end{figure*}

The wall-normal profiles of the dimensionless $\C P^{\dagger}=\C P \kappa/  (V^0\sb{ns})^2$  contributions to the production term are shown in \Fig{f:prod3}.   The main contribution $\C P_1$, shown by purple dotted lines, is peaking in the channel core, where it is almost constant, then quickly decreasing toward the walls. This behavior is very similar to $I_{\ell,x}(y)$ at all studied temperatures, with differences in the near-wall behavior. For  the parabolic $V\sb n$ profiles, \Fig{f:prod3}a-c,  at high temperature,  $\C P_1$ remains non-zero even very close to the walls, at intermediate $T=1.65$ K $\C P_1$ drops to zero at about intervortex distance from the wall, while at low $T$ it becomes negligible already at about $2 \ell$ from the nearest wall.   Two other contributions, $\C P_2$ and $\C P_3$ are negligible compared to $\C P_1$ in the channel core, gradually increasing toward the walls and attaining the largest values at the distance $\ell$ from them. Here we see the largest difference between the high and low $T$ behavior. At $T=1.9$ K, the  contributions of $\C P_2$ and $\C P_3$ may be safely neglected everywhere in the channel. At $T=1.65$ K the contribution of $\C P_2$  becomes important, while at $T=1.3$ K both $\C P_2$ and $\C P_3$ are dominant in near  the walls, such that overall production in this region is about half of that  in the channel core. As a results, the total production $y$-profile becomes similar to that for the flattened $V\sb n$ profile at this temperature, \Fig{f:prod3}d, although in the latter case $\C P_1$ has the dominant contribution (about 90\%) everywhere  in the channel. For this type of the $V\sb n$ profile, the contributions of $\C P_2$ and $\C P_3$ may be neglected  at all temperatures, especially at high $T$. The difference between the production in the channel core and in the near-wall regions is much smaller than for the parabolic $V\sb n$ profiles. These features are even more pronounced at higher temperatures. 

\begin{figure*}[htp]
  	\includegraphics[scale=0.45]{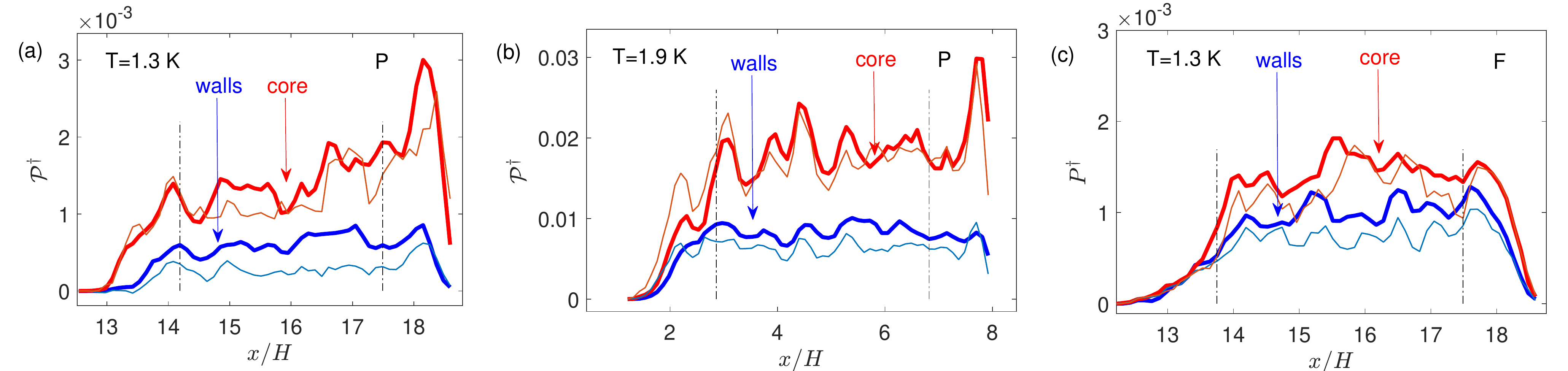}\\
  	 	\includegraphics[scale=0.45]{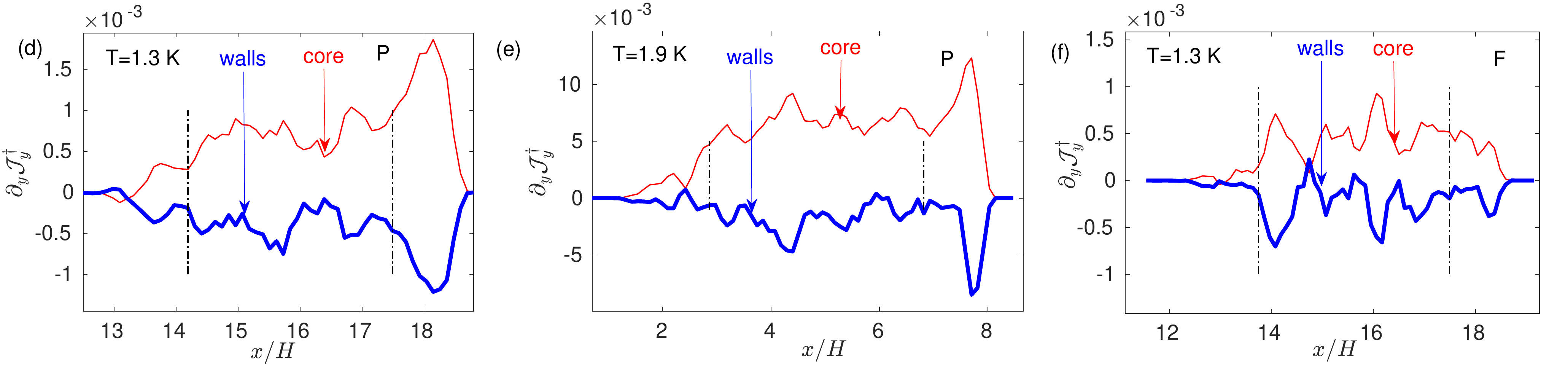}
  	\caption{\label{f:prodX} (a-c) The  streamwise profiles of the production terms:  $\C P_1$ (\Eq{p1}, thin lines),   and the total production $\C P_1+\C P_2+ \C P_3$ (thick  lines)   for different conditions.   (a) $T=1.3$~K, parabolic $V\sb n$ with $U_c=3$ cm/s, (b) $T=1.9$~K, parabolic $V\sb n$ with $U_c=1$ cm/s, (c) $T=1.3$~K, flattened $V\sb n$ profile.	 (d-f)  The  streamwise profiles of  $\partial \C J_y/\partial y$. To calculate the derivative,  at each $x$-point $\C J_y(y)$ was fitted by 7th-degree polynomial function.  The resulting $\partial \C J_y(x,y)/\partial y$ was averaged over the core and the walls regions. The profiles  $\partial \C J\sp{wall}_y/\partial y$ is a sum over both walls regions. Thin dot-dashed black lines mark the edges of the bulk  region.  }
  \end{figure*}

To see how the VLD production is distributed  along the tangle, we plot in \Fig{f:prodX}(a-c)
the streamwise profiles of $\C P^{\dagger}_1$ and of the total production $\C P^{\dagger}_1+\C P^{\dagger}_2+ \C P^{\dagger}_3$  for the same conditions as in \Fig{f:prod3}. We do not show the profiles for $T=1.65$ K, as they represent an intermediate case and do not bring more information. 

First of all, we can clearly distinguish the bulk, the hot and the cold front regions. In the  tangle bulk, the  production  is almost constant, up to fluctuations that are stronger in the channel core than in the near-walls region. In accordance with profiles shown in \Fig{f:prod3},  the contribution of $\C P_1$  (thin lines) is dominant at high $T$, \Fig{f:prodX}b, both in the core and near walls, as well as for the flows generated by the flattened $V\sb n$ profiles, \Fig{f:prodX}c. At low $T$, \Fig{f:prodX}a, $\C P_1$ constitutes about  half of the total production in the near-walls region. 

In the hot front region, the production in the core has a pronounced peak in the channel core, very close to the tangle edge,  which is dominated by $\C P_1$.  The VLD  produced in this region is then taken to the walls by the transverse flux, as is well seen in  \Fig{f:prodX}(d-f) where we plot  $\partial \C J^{\dagger}_y(x ,t)/\partial y,$ for the dimensionless $\C J_y^{\dagger}=\C J_y \kappa/  (V^0\sb{ns})^2$.  Although this peak is not as pronounced in the flows generated by the flattened $V\sb n$ profiles, the production is still stronger in the channel core   than near the walls.  The horseshoe shape of the VLD distribution, as in \Fig{f:10}c, is the result of this  dominant production in the channel core and the outward flux in the hot front region.

The situation is completely different in the cold front region, where the production  and the fluxes are strongly suppressed. Here, the production, the decay, and the fluxes balance each other in a  manner that strongly depend on the flow conditions.


\end{document}